\documentclass{jfm}

\usepackage{graphicx}
\usepackage{newtxtext}
\usepackage{newtxmath}
\usepackage{natbib}
\usepackage{amsmath}
\usepackage{pdflscape} 
\usepackage[figuresright]{rotating}
\usepackage{booktabs}  
\usepackage{siunitx}  
\usepackage{tabularx}
\usepackage{xcolor}
\usepackage{multirow}
\usepackage{hyperref}
\hypersetup{
    colorlinks = true,
    urlcolor   = blue,
    citecolor  = black,
}

\newcommand{\RomanNumeralCaps}[1]
\linenumbers

\title{Turbulent diffusivity and effective rise velocity of buoyant particles in a free-surface boundary layer}

\author{Julio E. Chávez-Dorado\aff{1},
Lucia J. Baker\aff{1},
James J. Riley\aff{1},
\and Michelle H. DiBenedetto\aff{1,2}\corresp{\email{mdiben@princeton.edu}}}

\affiliation{\aff{1}Department of Mechanical Engineering, University of Washington, Seattle, Washington, USA
\aff{2}Department of Mechanical and Aerospace Engineering, Princeton University, Princeton, New Jersey, USA}

\begin{document}
\maketitle

\newcommand{\waveamp}{a} 
\newcommand{\wavepeakamp}{a_p} 
\newcommand{\wavesigamp}{a_{s}} 
\newcommand{\majax}{a} 
\newcommand{\partRad}{a} 
\newcommand{\jonampspecwi}{a_i} 
\newcommand{\pr}{a_p} 
\newcommand{\dps}{A} 
\newcommand{\acorram}{A} 
\newcommand{\bestA}{\bm{\mathsf{A}}} 
\newcommand{\bestAtil}{\Tilde{\bm{\mathsf{A}}}} 

\newcommand{\minax}{b} 
\newcommand{\initconds}{\bm{\mathsf{b}}} 
\newcommand{\Bin}{\mathcal{B}} 
\newcommand{\bin}{b} 

\newcommand{\concInit}{c_0} 
\newcommand{\conc}{c} 
\newcommand{\concz}{c(z)} 
\newcommand{\concavg}{\langle c \rangle} 
\newcommand{\concf}{c'} 
\newcommand{\wavephsp}{c_p} 
\newcommand{\ctrajfac}{C} 
\newcommand{\dragcoef}{C_d} 
\newcommand{\amasscoeff}{C_m} 
\newcommand{\specC}{\mathbf{C}} 

\newcommand{\partDiam}{d_p} 
\newcommand{\massDiff}{D} 
\newcommand{\molDiff}{D} 


\newcommand{\wavefreq}{f} 
\newcommand{\schinau}{f_D} 
\newcommand{\wavepeakfreq}{f_p} 
\newcommand{\samfreq}{f_s} 
\newcommand{\concpdf}{f_Z} 
\newcommand{\Frp}{Fr} 

\newcommand{\gvec}{\mathbf{g}} 
\newcommand{\gmag}{g} 
\newcommand{\Ga}{Ga} 
\newcommand{\gfmag}{\gmag_f} 
\newcommand{\gfvec}{\gvec_f} 

\newcommand{\waterdepth}{h} 
\newcommand{\mixld}{h_{\text{ML}}} 
\newcommand{\sigwaveheight}{H_s} 
\newcommand{\hankmat}{\bm{\mathsf{H}}} 


\newcommand{\Jgerbi}{J_{33}} 

\newcommand{\wavenum}{k} 
\newcommand{\peakwavenum}{k_p} 
\newcommand{\jonwavenum}{k_i} 
\newcommand{\diffk}{K} 
\newcommand{\diffkf}{K_{f}} 
\newcommand{\diffkn}{K_{n}} 
\newcommand{\diffkzd}{K_{A}} 
\newcommand{\diffkze}{K_{C}} 
\newcommand{\diffkzm}{K_{F}} 
\newcommand{\diffkzt}{K_{T}} 
\newcommand{\diffkzv}{K_{V}} 
\newcommand{\diffks}{K^S} 

\newcommand{\charlp}{\ell} 
\newcommand{\kls}{\ell_\eta} 
\newcommand{\klse}{\hat{\ell}_\eta} 
\newcommand{\La}{La} 
\newcommand{\eullsc}{L_E} 
\newcommand{\mixlen}{\ell_m} 

\newcommand{\nrows}{m} 
\newcommand{\massf}{m_f} 
\newcommand{\massp}{m_p} 
\newcommand{\genmat}{\bm{\mathsf{M}}} 
\newcommand{\genmattrun}{\bm{\mathsf{M}}_r} 

\newcommand{\ncols}{n} 
\newcommand{\totelem}{N} 
\newcommand{\noisef}{N_f} 
\newcommand{\jonnumwave}{N_w} 


\newcommand{\prt}{\mathcal{P}} 
\newcommand{\arsym}{p_{\parallel}} 
\newcommand{\arper}{p_{\perp}} 

\newcommand{\turbpar}{Q} 
\newcommand{\turbparb}{Q_b} 
\newcommand{\turbpare}{\hat{Q}} 

\newcommand{\moder}{r} 
\newcommand{\acorr}{R} 
\newcommand{\Rep}{Re_p} 
\newcommand{\Ret}{Re_t} 
\newcommand{\Rew}{Re_w} 
\newcommand{\Ro}{Ro} 
\newcommand{\Sv}{Rv} 

\newcommand{\waveampspec}{S_{\zeta \zeta}} 
\newcommand{\wvelpspec}{S_{w'w'}} 
\newcommand{\Sc}{Sc} 
\newcommand{\Sct}{Sc_t} 
\newcommand{\SG}{SG} 
\newcommand{\St}{St} 
\newcommand{\Sz}{Sz} 
\newcommand{\tseg}{\mathcal{S}} 
\newcommand{\tlagor}{s_0} 

\newcommand{\tm}{t} 
\newcommand{\tsdur}{T} 
\newcommand{\lagtsc}{T_L} 
\newcommand{\lagtsclt}{T^{(a)}_L} 
\newcommand{\lagtscls}{T^{(t)}_L} 
\newcommand{\eultsc}{T_E} 
\newcommand{\lagtsct}{T'_L} 
\newcommand{\lagtscw}{\tilde{T}_{L}} 
\newcommand{\totT}{T} 
\newcommand{\wavepeakper}{T_p} 

\newcommand{\uVel}{u} 
\newcommand{\uavg}{\langle u \rangle} 
\newcommand{\uwave}{\Tilde{u}} 
\newcommand{\uf}{u'} 
\newcommand{\ufp}{\mathbf{u}_f} 
\newcommand{\up}{\mathbf{u}_p} 
\newcommand{\shearVel}{u_*} 
\newcommand{\shearVela}{u_{*\mathrm{a}}} 
\newcommand{\shearVelw}{u_{*\mathrm{w}}} 
\newcommand{\uworb}{u_\text{orb}} 
\newcommand{\uwaverms}{\tilde{u}_{\text{rms}}} 
\newcommand{\slipvelmag}{u_s} 
\newcommand{\uturbrms}{u_{\text{rms}}} 
\newcommand{\Uten}{U_{10}} 
\newcommand{\jonwavelamp}{U_i} 
\newcommand{\velf}{\mathbf{u}_f} 
\newcommand{\singleftvec}{\mathbf{u}_k} 
\newcommand{\velp}{\mathbf{u}_p} 
\newcommand{\vels}{\mathbf{u}_s} 
\newcommand{\singleft}{\bm{\mathsf{U}}} 
\newcommand{\velSto}{U_S} 

\newcommand{\vVel}{v} 
\newcommand{\riseVelq}{v_q} 
\newcommand{\singrightvec}{\mathbf{v}_k} 
\newcommand{\singright}{\bm{\mathsf{V}}} 

\newcommand{\wVel}{w} 
\newcommand{\wavg}{\langle w \rangle} 
\newcommand{\wwave}{\Tilde{w}} 
\newcommand{\wf}{w'} 
\newcommand{\wvelf}{w_f} 
\newcommand{\wvelp}{w_p} 
\newcommand{\wvelq}{w_0} 
\newcommand{\wvels}{w_s} 
\newcommand{\driftVel}{w_d} 
\newcommand{\riseVel}{w_r} 
\newcommand{\wturbrms}{w_{\text{rms}}} 
\newcommand{\wturbrmse}{\hat{w}_{\text{rms}}} 
\newcommand{\atilevecs}{\bm{\mathsf{W}}} 
\newcommand{\wplagr}{{w_p}} 
\newcommand{\wplagrf}{{w_p'}} 
\newcommand{\wplagrInit}{W^{+}_0} 
\newcommand{\wflagr}{w_f} 
\newcommand{\wflagrf}{w_f'} 
\newcommand{\wflagrInit}{W_0} 

\newcommand{\xDir}{x} 
\newcommand{\xp}{\mathbf{x}_p} 
\newcommand{\xInit}{x_0} 
\newcommand{\partDisp}{X} 
\newcommand{\colsnap}{\mathbf{x}_k} 
\newcommand{\presst}{\bm{\mathsf{X}}} 
\newcommand{\futst}{\bm{\mathsf{X}}'} 
\newcommand{\pseudoinv}{\bm{\mathsf{X}}^+} 

\newcommand{\yDir}{y} 

\newcommand{\zDir}{z} 
\newcommand{\zInit}{z_0} 
\newcommand{\turbroughsc}{z_0} 
\newcommand{\RoSed}{Z} 
\newcommand{\zplagr}{{z_p}} 
\newcommand{\vertdispvar}{\langle \Delta \zplagr^2(\tlag) \rangle} 
\newcommand{\zplagrInit}{Z^{+}_0} 
\newcommand{\zflagr}{Z^+} 
\newcommand{\zflagrInit}{Z_0} 

\newcommand{\jonalpha}{\alpha} 
\newcommand{\geralpha}{\alpha} 

\newcommand{\jonbeta}{\beta} 
\newcommand{\denpar}{\beta} 

\newcommand{\evc}{\gamma} 
\newcommand{\resrate}{\gamma} 

\newcommand{\waved}{\delta} 

\newcommand{\normerr}{\epsilon} 
\newcommand{\rdendiff}{\epsilon} 
\newcommand{\dissturb}{\varepsilon} 
\newcommand{\dissturbe}{\hat{\varepsilon}} 

\newcommand{\wfreesurf}{\zeta} 


\newcommand{\eflc}{\theta} 


\newcommand{\vonkarc}{\kappa} 

\newcommand{\dmdeval}{\lambda} 
\newcommand{\wavelen}{\lambda} 
\newcommand{\peakwavelen}{\lambda_p} 
\newcommand{\aevals}{\bm{\mathsf{\Lambda}}} 

\newcommand{\viscd}{\mu} 

\newcommand{\visck}{\nu} 
\newcommand{\eddyVisc}{\nu_t} 
\newcommand{\eddyViscs}{\nu^S_t} 

\newcommand{\tLag}{\xi} 
\newcommand{\acorrer}{\xi} 



\newcommand{\den}{\rho} 
\newcommand{\denf}{\rho_{f}} 
\newcommand{\pcorr}{\rho_{P}} 
\newcommand{\denp}{\rho_{p}} 
\newcommand{\acorru}{\rho_u} 
\newcommand{\acorrw}{\rho_{\wf \wf}} 

\newcommand{\waveage}{\sigma} 
\newcommand{\singvalel}{\sigma_k} 
\newcommand{\singval}{\bm{\mathsf{\Sigma}}} 

\newcommand{\shstr}{\tau} 
\newcommand{\dvar}{\tau} 
\newcommand{\tlag}{\tau} 
\newcommand{\acorrdt}{\tau'} 
\newcommand{\acorrdw}{\Tilde{\tau}} 
\newcommand{\rtimef}{\tau_{\eta}} 
\newcommand{\rtimefe}{\hat{\tau}_{\eta}} 
\newcommand{\rtimep}{\tau_p} 

\newcommand{\volfrac}{\upsilon} 

\newcommand{\dmdmode}{\phi} 
\newcommand{\wavephase}{\phi} 
\newcommand{\axar}{\Phi} 
\newcommand{\jonwavephase}{\phi_i} 
\newcommand{\aevecs}{\bm{\mathsf{\Phi}}} 
\newcommand{\stabfunmo}{\varphi} 


\newcommand{\wavedmdmodes}{\psi_{\text{wave}}} 
\newcommand{\relwaverank}{\psi_{\text{wave}}/r} 

\newcommand{\jonangfreq}{\omega} 
\newcommand{\jonangfreqpeak}{\omega_p} 
\newcommand{\acontevals}{\bm{\mathsf{\Omega}}} 

\begin{abstract}
Predicting the transport of buoyant particles in a free-surface boundary layer is important to the study of many environmental systems, including microplastics in the upper ocean. 
Current transport models, adapted from sediment transport theory, typically rely on assumptions of a quiescent rise velocity and gradient diffusion with an uncertain turbulent Schmidt number $\Sct$. Here, we test this type of model against experiments by studying the vertical mixing of near-neutrally buoyant, finite-size spheres, rods, and disks in a wind-driven, wavy free-surface flow. We measure particle diffusivity directly from Lagrangian trajectories and compare against Eulerian concentration-based estimates. 
Overall, we find that particle buoyancy is the main control on the diffusivity, and that the diffusivity decreases as particle rise velocity grows relative to the turbulent fluctuations. These observations we find to be consistent with the crossing-trajectories theory, even in the presence of waves.
In addition, we find that inferring the diffusivity from concentration profiles with an assumed quiescent rise velocity overestimates the diffusivity by up to a factor of $5$, consistent with effective rise velocities up to $80\%$ lower than the corresponding quiescent values. We also directly measure $\Sct\approx1$ for the neutrally-buoyant particles and $\Sct>1$ for the buoyant particles. Together, these experimental results demonstrate how standard model closures may be biased in both their diffusivities and rise velocities when applied to buoyant particles at the ocean surface.  
\end{abstract}

\begin{keywords}
must be chosen during online submission (see \href{https://www.cambridge.org/core/journals/journal-of-fluid-mechanics/information/list-of-keywords}{Keyword PDF} for full list). 
\end{keywords}

\section{Introduction}
\label{sec:intro}
Particle transport in boundary layer flows is common in the environment, from sediment transport along river beds to bubbles and plastics at the ocean surface. 
While sediment transport in a bottom boundary layer is a well-studied system, the transport of buoyant particles in a wind-driven, free-surface boundary layer has received comparatively less attention. 
Even less studied are near-neutrally buoyant, but finite-size particles such as microplastics \citep{DiBenedetto_2026} and oil droplets \citep{Boufadel_2020}.
Understanding the vertical transport of these particles is especially important because it regulates shear-induced horizontal dispersion \citep{Laxague_2018} and determines particle residence times at the surface, controlling exposure to sunlight and thereby photodegradation \citep{Ward_2019}.

Predicting the vertical transport of buoyant particles typically requires accurate models for two quantities: the particles' turbulent diffusivity and their effective rise velocity. 
With respect to diffusivity, established models exist for diffusivity profiles throughout the mixed layer, but the near-surface wave layer remains challenging due to wave orbital motion, wave breaking, and Langmuir circulation \citep{Qiao_2016, Gerbi_2009, DAsaro_2014}. 
Additionally, the diffusivity of finite-size particles is not necessarily the same as that of fluid tracers. 
This difference is quantified by the turbulent Schmidt number $\Sct=\eddyVisc/\diffk$, the ratio of the eddy viscosity to the particle diffusivity.  
In the sediment-transport literature, contrasting studies have reported both $\Sct>1$ and $\Sct<1$ \citep{Gualtieri_2017}.  
This discrepancy can partly be attributed to the ways in which diffusivity is inferred from measurements. 
For example, when the diffusivity is indirectly measured from concentration profiles by assuming a still-water settling velocity, diffusivity can be over-estimated \citep{Chauchat_2022, Li_2023}. 
When the diffusivity is measured via direct measurements of the turbulent particle flux, one finds lower diffusivities and $\Sct = 3-4$ for settling sand \citep{Chauchat_2022}. 
This study also implies that the effective particle settling velocity is reduced relative to the quiescent value. 
Thus, the diffusivity must be measured without assuming a settling (or rise) velocity.

With respect to the rise velocity, the quiescent rise velocity is not necessarily the same as the effective rise velocity of the particles in a flow. 
Several mechanisms can modify it. For example, buoyant particles can preferentially sample downwelling parts of the flow, either kinematically, by collecting in zones of surface convergence above downwelling \citep{Stommel_1949, Chor_2018a, Yang_2014} or dynamically, through inertial vortex trapping \citep{Aliseda_2011}. 
Nonlinear drag can also reduce the drift of particles with finite particle Reynolds number \citep{Good_2014,Ruth_2021}, unsteady forces can reduce it further for particles larger than the Kolmogorov scale \citep{Li_2023}, and waves can modulate it directly \citep{Clark_2020, DiBenedetto_2022}. 
Large-eddy simulations of Langmuir turbulence with heavy non-inertial particles have shown reduced settling velocities due to the coherent structures \citep{Noh_2006, Chamecki_2019}. 
Thus, assuming a quiescent rise velocity may be inaccurate at the ocean surface. 

For buoyant particles in steady state, the vertical concentration profile is set by a balance between upward advection due to the particles rising and downward turbulent flux due to mixing \citep{Kukulka_2012}. 
A measured concentration profile therefore constrains only the ratio of rise velocity to the diffusivity, and any error in assumed rise velocity will map directly onto the inferred diffusivity (and vice versa). 
Transport by large coherent structures (e.g., Langmuir circulations) can enter this balance either as additional mixing via a nonlocal flux or as a change in the effective rise velocity \citep[e.g., ][]{Smyth_2002, Giordani_2020}. 
Thus,  measurements of only concentration profiles can obscure the relevant processes affecting the particles; this motivates our use of particle tracking in our experiments.

The diffusivity of particles is modulated by their interactions with the flow. 
For example, particles with a net drift (e.g. settling or rising due to gravity) experience the \emph{crossing-trajectories} effect: they cross fluid pathlines as they drift causing their turbulent velocity fluctuations to decorrelate faster than that of a tracer particle, reducing their diffusivity even in the absence of particle inertia \citep{Csanady_1963}. 
The drift velocity relative to the turbulent fluctuations, here defined as $\Sv$ (where $\Sv$ stands for rise velocity here, but is identical to the $Sv$ number commonly used to describe a non-dimensional settling velocity), is therefore a controlling parameter for the particle diffusivity. 
In contrast,  particle inertia filters rapid fluctuations in the flow, which can increase the particle's velocity decorrelation time and thereby enhance the particle's diffusivity relative to a tracer \citep{Reeks_1977, Squires_1991, Wetchagarun_2010}. 
These effects are commonly characterized by the Stokes number $\St=\rtimep/\rtimef$, the ratio of the particle response time to a characteristic fluid timescale, and the size ratio $\Sz=\partDiam/\kls$,  the ratio of the particle size to the characteristic flow length scale.
Most experiments on particle diffusivity mainly concern small, heavy particles in homogeneous isotropic turbulence \citep{Balachandar_2010, Brandt_2022}. 
Near-neutrally buoyant, finite-size particles are outside this regime; their density ratio makes inertial filtering weak \citep{Berk_2024}, but their size means they will filter the flow spatially \citep{Qureshi_2007}. 
With respect to buoyant particles,  buoyant droplets in isotropic turbulence have shown reduced diffusivity consistent with the crossing-trajectories effect \citep{Gopalan_2008}, and reduced rise velocities for large droplets \citep{Friedman_2002}, while finite-size floating particles in free-surface turbulence showed reduced velocity fluctuations but longer correlation times, leading to only a weakly size-dependent diffusivity \citep{Salmon_2025}. 
However, it remains to be seen how these results translate to buoyant particles in a wavy, turbulent flow.

In practice, particle diffusivity is estimated in one of two ways. Lagrangian methods follow the particles; e.g., \citet{Taylor_1922} relates the long‑time (asymptotic) growth of particle displacements to the velocity correlation along their trajectories. 
This framework provides a direct measure of dispersion with no assumptions about the particle drift.
However, this method requires measuring long trajectories to achieve converged statistics, measurements that are difficult to obtain in the field. 
Eulerian measurements instead interpret  concentration profiles through the flux balance above, following sediment transport theory \citep{Rouse_1937}.
This approach has been adapted for microplastics at the ocean surface \citep{Kukulka_2012}, where the near-surface diffusivity is often parametrized from wind and wave conditions or from large-eddy simulations, and the drift velocity is prescribed as the quiescent rise velocity \citep{Kukulka_2015, Chor_2018, Chamecki_2019}. 
These inherently Eulerian models are widely used to interpret microplastics observations, but they have yet to be verified in the lab. Thus, we aim to study how the assumptions associated with these Eulerian closures compare against direct Lagrangian measurements.

In this study, we measure the vertical turbulent diffusivity of near-neutrally buoyant particles in a laboratory wind-driven, free-surface boundary layer directly from particle trajectories. 
We focus on the regime where particle rise velocity and turbulent fluctuations are similar in magnitude ($\Sv\sim 1$), so that the particles are able to partially mix below the surface. 
We find that particle buoyancy is the primary control on the diffusivity, and that the crossing-trajectories theory of \citet{Csanady_1963} describes our measurements well, even for finite-size and non-spherical particles in the presence of waves. In contrast, we find that the quiescent rise velocity is a poor descriptor of the mean advective transport, and that the effective rise velocity is reduced far below the quiescent value. 
The remainder of the paper is structured as follows: Section \ref{sec:Background} outlines Lagrangian and Eulerian diffusivity estimation methods; Section \ref{sec:Methodology} details the experimental setup, flow conditions, and particle properties; Section \ref{sec:Results} presents and discusses the findings; and Section \ref{sec:Conclusions} summarizes the key results and their transport modeling implications.

\section{Background}
\label{sec:Background} 
In this section, we review the approaches that we use to estimate particle turbulent diffusivity in our experiments, summarized at the end of this section in Table \ref{tab:methods}.
The Lagrangian approach uses particle trajectories to estimate the diffusivity from the particle mean-square displacement (MSD).
The Eulerian approaches use concentration profiles to estimate the diffusivity from the one-dimensional flux balance.
We note that, while a true \emph{Lagrangian} (or \emph{fluid}) particle moves exactly with the flow and thus behaves as a tracer, here we use the term Lagrangian more loosely to refer to quantities defined along a particle trajectory, whether the particle is a tracer or not.
Throughout, we assume the flow is statistically stationary and horizontally homogeneous, and we focus on the vertical dynamics with $\zDir$ positive upward and $\zDir=0$ defined at the mean water level.

First, we define the relevant particle velocities we refer to below. 
The quiescent rise velocity $\wvelq$ is the particle's terminal velocity in still water. 
The drift velocity $\riseVel$ is the mean velocity of the particles relative to the mean velocity of the fluid; it is commonly assumed that $\riseVel = \wvelq$. For our particles, the effective rise velocity $\riseVel$ is the drift velocity we infer from our measurements (\S\ref{subsec:effrisevel}). Finally, $\langle \wplagr \rangle$ is the measured ensemble-mean particle velocity, which vanishes in a statistically steady, vertically bounded system (\S\ref{subsec:Asymptotic}).
In the following, we consider the different limiting cases that can influence the measured particle diffusivity. 
We first derive the standard Lagrangian diffusivity for particles with zero mean vertical velocity, and then consider the more general case of particles with a nonzero mean velocity. 
 
\subsection{(Lagrangian) particle trajectory approach}
\label{subsec:Lagrangian_methds}
Dispersion is an inherently Lagrangian phenomenon. 
For a given particle, we denote its instantaneous vertical position by $\zplagr(\tm)$, and its displacement as 
\begin{equation}
    \Delta \zplagr(\tm) = \zplagr(\tm +\tm_0) - \zplagr(\tm_0),
\end{equation}
where $\zplagr(
\tm_0)$ is the initial position of the particle at time $\tm_0$.  The most direct measure of the spread of an ensemble of particles comes from the long-time growth of the particle MSD, i.e., dispersion.

\citet{Einstein_1905} showed that the spread of a cloud of particles with uncorrelated motion, as measured by the MSD $\langle \Delta \zplagr^2(\tm) \rangle$, will grow linearly with time for sufficiently long times: 
\begin{equation}
    \langle \Delta \zplagr^2(\tm) \rangle \sim 2\diffk \tm,
\end{equation}
where $\diffk$ is the particle diffusivity and $\langle \cdot \rangle$ denotes an ensemble average.
\citet{Taylor_1922} extended this to particles with finitely correlated motion. To demonstrate, we define the displacement as the time-integral of the particle's vertical velocity $\wplagr(\tm)$:
\begin{equation}
    \Delta \zplagr(\tm) = \int_0^{\tm} \wplagr(\tm') \, d\tm'.
    \label{eqn:diff_kin}
\end{equation}
Next, we decompose the  velocity into the mean and fluctuation $\wplagr(\tm) = \langle \wplagr \rangle + \wplagrf(\tm)$.  
For a statistically stationary flow with no mean vertical particle velocity $(\langle \wplagr \rangle = 0)$, the MSD can be expressed in terms of the velocity autocovariance $\langle \wplagrf(\tm)\wplagrf(\tm + \tlag)\rangle$:
\begin{equation}
    \langle \Delta \zplagr^2(\tm) \rangle =  2\int_0^\tm (\tm - \tlag)\langle \wplagrf(\tm)\wplagrf(\tm + \tlag)\rangle\, d\tlag.
    \label{eqn:diffk_msd}
\end{equation}
We note that even in flows with zero mean, 
finite-size particles can preferentially sample the flow, which can yield a non-zero mean drift $\langle \wplagr \rangle \neq 0$ in an unbounded flow.
We will revisit these implications later in this section.

We can relate the MSD to the Lagrangian integral time scale $\lagtsc$ which is defined as
\begin{equation}
    \lagtsc=\int_0^\infty \acorrw \, d\tlag, \quad \text{where} \quad \acorrw = \frac{\langle \wplagrf(\tm)\wplagrf(\tm + \tlag)\rangle}{\langle \wplagrf^2\rangle}
    \label{eqn:diffk_its}
\end{equation}
is the Lagrangian velocity autocorrelation. From this expression we see the two classic limiting regimes for \eqref{eqn:diffk_msd}: the \emph{ballistic} and the \emph{diffusive} regimes. 
In the ballistic regime ($\tm \ll \lagtsc$), the velocity remains correlated $(\acorrw \approx1)$, and the MSD grows quadratically in time,
$    \langle \Delta \zplagr^{2}\rangle\approx 2\langle \wplagrf^2\rangle\,\tm^2.$ 
In the diffusive ($\tm\gg \lagtsc$) regime, the velocity fluctuations become uncorrelated and the MSD grows linearly in time, $\langle \Delta \zplagr^{2}\rangle\approx 2\langle \wplagrf^2\rangle\,\lagtsc\tm.$ The time derivative of the MSD thus approaches a constant over sufficiently long times:
\begin{equation}
    \lim_{\tm\to \infty} \frac{d}{d\tm}\langle \Delta \zplagr^{2}(\tm)\rangle = \int_0^\infty \langle \wplagrf(\tm)\wplagrf(\tm + \tlag)\rangle\, d\tlag = 2 \diffk.
    \label{eqn:difftdir}
\end{equation}
This yields an asymptotic definition of turbulent (vertical) diffusivity based on particle dispersion:
\begin{equation}
    \diffkzd = \lim_{\tm \to \infty} \frac{1}{2} \frac{d}{d\tm}\langle \Delta \zplagr^{2}(\tm)\rangle.
    \label{eqn:difft}
\end{equation}
Equivalently, from \eqref{eqn:diffk_its} and \eqref{eqn:difftdir}, the diffusivity can be expressed in terms of the Lagrangian integral time scale as
\begin{equation}
 \diffk = \int_0^\infty \langle \wplagrf(\tm)\wplagrf(\tm + \tlag)\rangle\, d\tlag = \langle \wplagrf^2\rangle \lagtsc.
 \label{eqn:diffk_taylor}
\end{equation}
Equations \ref{eqn:difft} and \ref{eqn:diffk_taylor} are equivalent definitions of the same diffusivity, assuming statistical stationarity (the autocorrelation depends only on the time lag $\tlag$), homogeneity (a single $\diffk$ characterizes the region sampled by the particle trajectories), and the existence of a diffusive regime. These assumptions can break down in boundary layers where $\acorrw$ and $\lagtsc$ can vary with depth.

\subsubsection{Effects of drift velocity}
\label{sec:inertia_crosstraj}
Next, we consider turbulent diffusivity in the context of particles with a mean drift. Non-fluid particles can have a mean vertical drift velocity associated with gravity. Alternatively, the flow may have  a non-zero mean flow. In either case, equation \ref{eqn:diff_kin} becomes:
\begin{equation*}
\Delta \zplagr(\tm) = \langle \wplagr\rangle \tm + \int_{0}^{\tm}\wplagrf(\tm')\,d\tm',
\end{equation*}
where $\langle \wplagr\rangle $ is some constant, consistent with the assumptions of homogeneity and stationarity above.
Therefore, \eqref{eqn:diffk_msd} generalizes to:
\begin{equation}
\langle \Delta \zplagr^2(\tm) \rangle = \langle \wplagr\rangle^2 \tm^2 + 2\int_0^\tm (\tm - \tlag)\langle \wplagrf(\tm)\wplagrf(\tm + \tlag)\rangle\, d\tlag, \label{eqn:tay16}
\end{equation}
and the diffusive limit is
\begin{equation}
\frac{1}{2}\frac{d}{d \tm}\langle \Delta \zplagr^2(\tm) \rangle = \langle \wplagr\rangle^2 \tm + \int_0^\tm \langle \wplagrf(\tm)\wplagrf(\tm')\rangle \,d\tm'. \label{eqn:tay17}
\end{equation}
The drift term $\langle \wplagr\rangle^2 \tm$ represents the ``ballistic" contribution to the variance of the particle position relative to the origin, caused by the non-zero mean drift and/or a non-zero mean flow.
If the MSD is calculated relative to a frame of reference moving with the drift, this term vanishes and we recover the Taylor result.

There are now two conditions to be met for \eqref{eqn:difft} to be used to compute the diffusivity $\diffk$. 
The first is that time $\tm$ is large enough that the integral in \eqref{eqn:tay17} has become a constant,
i.e., approximately, $\tm > \lagtsc$, and the second is that time is short enough so that the first term in
\eqref{eqn:tay17} does not dominate the second, i.e., approximately $ \langle \wplagrf^2\rangle \lagtsc > \langle \wplagr\rangle^2 \tm$. 
Taken together, these conditions give:
\begin{equation}
    \frac{\langle \wplagrf^2\rangle}{\langle \wplagr\rangle^2}\lagtsc > \tm > \lagtsc,
\end{equation}
or, since $\tm > \lagtsc$ is necessary,
\begin{equation}
    \frac{\langle \wplagrf^2\rangle}{\langle \wplagr\rangle^2} > 1.
    \label{eqn:wrmswr}
\end{equation}
We evaluate this condition for our experiments in \S\ref{subsec:Asymptotic}. 

Additionally, the crossing-trajectories effect becomes important when particles experience a net drift velocity relative to the fluid \citep{Yudine_1959, Csanady_1963}. 
This effect is governed by the velocity ratio
\begin{equation}
    \Sv = \frac{\wvelq}{\upsilon_f},
\end{equation}
where $\upsilon_f$ is a characteristic velocity scale of the flow (in our experiments, we use $\upsilon_f=\wturbrms$, the turbulent vertical velocity fluctuation root-mean-square).
A finite drift velocity causes a particle to travel through an eddy more quickly than a fluid element, reducing its correlation time and hence the diffusivity.
\citet{Csanady_1963} showed that, for dispersion in the drift direction, the crossing-trajectories effect can be captured by a ratio of particle-to-fluid diffusivity $\diffk/\diffkf$:
\begin{equation}
    \frac{\diffk}{\diffkf} = \left ( 1 + \frac{\ctrajfac^2 \wvelq^2}{\langle \wflagrf^2 \rangle} \right )^{-1/2}.
    \label{eqn:crosstraj}
\end{equation}
Here, $\ctrajfac$ can be interpreted as the ratio between the Lagrangian memory time of the fluid and the time required for drift to carry a particle across an Eulerian turbulent structure.
It follows that a larger $\ctrajfac$ produces a stronger reduction in diffusivity. 
The constant $\ctrajfac$ is not universal and  reported values ($\ctrajfac = 0.3\text{--}22$) span orders of magnitude across flows \citep[see][for specific values]{Snyder_1971, Wells_1983, Sato_1987, Burnage_1990, Mazzitelli_2004, Aiyer_2022}.

\subsubsection{Effects of inertia and finite-size}
\label{sec:finite_size}
In addition to drift, particle inertia and finite size can modify diffusivity by changing the two parameters in \eqref{eqn:diffk_taylor}. 
Inertia and finite size can filter high-frequency fluctuations, reducing the velocity variance $\langle \wplagrf^2\rangle $ while increasing the correlation time $\lagtsc$, so the net effect on $\diffk$ can be of either sign.

In homogeneous isotropic turbulence, inertia alone can increase the diffusivity of heavy particles at low Stokes numbers \citep{Reeks_1977}, with a weak decrease for $\St\gtrsim 1$ \citep{Wetchagarun_2010}.
\citet{Nir_1979} extended the analysis by \citet{Csanady_1963} and \citet{Reeks_1977} to include both a steady drift and inertia (via a linear drag), showing that the crossing-trajectories effect dominates, reducing the vertical diffusivity.
 A further extension showed that nonlinear drag can further reduce the drift and diffusivity \citep{Mei_1997}.
These results are generalized to arbitrary inertia, although no closed-form expression exists \citep{Boi_2018}.

For near-neutrally buoyant particles, however, inertial filtering is expected to be weak regardless of the response time. 
As the density ratio approaches unity, added-mass and pressure-gradient forces become relatively large and therefore a small particle's frequency response approaches that of a tracer \citep{Zhang_2019, Berk_2024}. 
Considering finite size,  a particle of size $\partDiam$ averages the fluid motion over its volume and filters eddies smaller than itself, a spatial rather than temporal filtering \citep{Qureshi_2007, Bec_2007, Toschi_2009}. 
We therefore expect drift, rather than inertia, to dominate the modulation of diffusivity in our regime, with finite size potentially acting as a secondary effect. 
We define the corresponding dimensionless parameters ($\St$, $\Sz$) with the appropriate flow scales for our finite-size particles in \S\ref{subsec:particles}.

\subsection{(Eulerian) concentration‐profile methods}
\label{subsec:concentration}
In practice, sediments and other particulate phases are typically modeled with an Eulerian conservation equation. 
Neglecting molecular diffusion (large P\'eclet number), the ensemble-averaged conservation equation for the vertical concentration of an ensemble of particles $\conc(\zDir,\tm)$ is
\begin{equation}
    \frac{\partial \concavg}{\partial \tm} + \frac{\partial}{\partial \zDir} \big(\langle \wplagr (\zDir) \rangle \concavg + \langle \wplagrf \conc'\rangle\big) = 0, \label{eq:c_reynolds}
\end{equation}
where $\conc=\concavg+\conc'$ and $\wplagr(\zDir)=\langle \wplagr (\zDir) \rangle+\wplagrf$ are the Reynolds decompositions of concentration and vertical particle velocity.
Note that here the mean vertical particle velocity $\langle \wplagr (\zDir) \rangle$ is an Eulerian ensemble average (different from the Lagrangian average $\langle \wplagr \rangle$) and accounts for the combined effects of the mean fluid velocity  and the particle drift, $\langle \wplagr (\zDir) \rangle=\langle \wflagr \rangle+\riseVel$.

Next, we parametrize the turbulent flux $\langle \wplagrf \conc'\rangle$ via the gradient-diffusion closure 
\begin{equation}
    \langle \wplagrf \conc'\rangle = -\diffk\frac{\partial\concavg}{\partial \zDir},
    \label{eqn:diff_graddiff}
\end{equation}
assuming that the turbulent flux is proportional to the local concentration gradient.
This, however, might not hold if larger scale motions (e.g., Langmuir circulation) exist, which can contribute to nonlocal vertical mixing \citep{Farmer_1994}.
Assuming zero mean fluid velocity ($\langle \wflagr \rangle=0$) and constant drift $\riseVel$, we have
\begin{equation}
\frac{\partial \concavg}{\partial \tm} + \riseVel \frac{\partial \concavg}{\partial \zDir}
= \frac{\partial}{\partial \zDir}\left(\diffk\,\frac{\partial \concavg}{\partial \zDir}\right),
\label{eq:fp}
\end{equation}
where $\diffk(\zDir, \tm)$ is an effective vertical eddy diffusivity.

We note that the Eulerian description above and the Lagrangian MSD approach are two representations of the same transport process (see Appendix \ref{app:equivalenceMSDconc}).  
The diffusivity in \eqref{eq:fp} equals the long-time growth rate of the MSD \citep{Majda_1999} for a cloud of dispersing particles. This equivalence is related to the assumptions of gradient diffusion and the existence of a diffusive regime. 
When these assumption hold and the drift velocity is known, the Lagrangian and Eulerian estimates should agree. 
Disagreement is therefore due to a failure of the underlying assumptions or an error in the drift velocity, rather than a difference in how the diffusivities are defined. We will compare these approaches in \S\ref{sec:Results}.

\subsubsection{Eulerian approaches to diffusivity}
Next, we assume steady state, such that equation \ref{eq:fp} becomes
\begin{equation}
\label{eq:vertDist_diff}
\riseVel \frac{\partial \concavg}{\partial \zDir} = \frac{\partial}{\partial \zDir} \left ( \diffk \frac{\partial \concavg}{\partial \zDir} \right ).
\end{equation} 
Given the condition of no net flux through the boundaries, the vertical flux throughout the water column is zero:
\begin{equation}
    0 = \riseVel \concavg - \diffk \frac{\mathrm{d}\concavg}{\mathrm{d}\zDir}.
    \label{eqn:diff_flux}
\end{equation}
Now, we can use equation \ref{eqn:diff_flux} to estimate diffusivity $\diffkzm$, which we refer to as the flux-estimate: 
\begin{equation}
    \diffkzm = \frac{\riseVel \concavg}{\mathrm{d}\concavg/\mathrm{d}\zDir}.
    \label{eqn:diff_f}
\end{equation}
This diffusivity $\diffkzm$ can be calculated directly from the concentration profile over depth, given $\riseVel$.
When $\riseVel$ is not known, the quiescent rise velocity is typically used.

The steady-state concentration profile is found by integrating \eqref{eqn:diff_flux},
\begin{equation}
\concavg (\zDir) = \concInit \exp \left ( \int^\zDir_0 \frac{\riseVel}{\diffk}\, d\zDir \right ).
\end{equation}
Further assuming that $\riseVel$ and $\diffk$ are constant over depth (i.e. $\riseVel = \wvelq$) yields the closed form solution 
\begin{equation}
    \concavg(\zDir) = \concInit \exp\left(\frac{\wvelq \zDir}{\diffk}\right),
\label{eq:vertDistributionModelKukulka}
\end{equation}
where $\concInit$ is the surface concentration.
Equation \ref{eq:vertDistributionModelKukulka} describes a profile with maximum concentration of particles at the surface decaying with depth  at the rate $\wvelq/\diffk$, which defines a single mixing length scale $\mixlen$ 
\begin{equation}
    \mixlen = \frac{\diffk}{\wvelq},
    \label{eqn:mixlen}
\end{equation}
characterizing the vertical extent over which the particles are mixed. If we fit an exponential profile to the particle concentration and know $\wvelq$, we have another method to estimate the turbulent diffusivity
\begin{equation}
    \diffkze = \mixlen\,\wvelq.
    \label{eqn:diff_mixlen}
\end{equation}
We refer to this as the concentration-profile estimate. 
In principle, another way to directly estimate  $\diffk$ is from inverting the gradient‑diffusion closure \eqref{eqn:diff_graddiff} using the measured turbulent flux $\langle \wplagrf \conc'\rangle$ rather than the mean flux.
However, $\langle \wplagrf \conc'\rangle$ can be challenging to measure because it requires co-located, simultaneous measurements of velocity  and concentration fluctuations, with sufficient averaging to reduce noise; we do not attempt it here. 

Finally, the particle diffusivity is often parameterized from the momentum transport using the Reynolds analogy, which relates $\diffk$ to the eddy viscosity $\eddyVisc$  via the turbulent Schmidt number
$\Sct=\eddyVisc/\diffk$, such that
\begin{equation}
    \diffkzv = \frac{\eddyVisc}{\Sct},
\end{equation}
where, in the absence of a measured or modeled value, $\Sct$ is typically set to unity assuming that turbulence tends to mix all things equally. This approach is useful in sediment transport because well-established turbulence models already exist to parametrize the eddy viscosity in a solid-wall boundary layer.
In our study of a free-surface boundary layer, we can estimate $\eddyVisc$ directly from the momentum flux and the mean shear as
\begin{equation}
    \eddyVisc = - \frac{\langle \uf \wf \rangle}{\partial \uavg/\partial \zDir}.
    \label{eqn:eddyviscgrad}
\end{equation}
While this direct estimation avoids empirical modeling biases, it assumes that the turbulent flux is driven by the local gradient, and it is susceptible to singularities where the shear vanishes.

\subsection{Summary}

\begin{table}
  \centering
  \caption{Summary of core formulas, principles, and input data to estimate vertical turbulent diffusivity $\diffk$ in our experiments.
}
  \begin{tabularx}{\linewidth}{@{} p{2.5cm}  
                                p{4.4cm}  
                                p{4.4cm}  
                                p{2cm}  
                                @{}}
    \toprule
    Formula          & Principle         & Assumptions           & Input Data       \\
    \midrule
    $\diffkzd =\frac{1}{2}\,\frac{\mathrm{d}\langle \Delta \zplagr^2 \rangle}{\mathrm{d}\tlag}$   & \emph{Asymptotic}: Long-time growth rate of particle mean-square displacement  & Homogeneous, stationary turbulence; $\langle \wplagrf^2\rangle/\langle \wplagr\rangle^2 > 1$; diffusive regime for $\tm \gg \lagtsc$  & $\zplagr(\tm)$
        \\
    \midrule
    $\diffkzm = \frac{\riseVel\,\concavg}{\mathrm{d}\concavg/\mathrm{d}\zDir}$ & \emph{Flux}: Balance between local buoyant flux to the local turbulent gradient flux & Steady state; local equilibrium (flux proportional to gradient)   & $\concavg(\zDir)$; $\mathrm{d}\concavg/\mathrm{d}\zDir$; $\riseVel$
       \\
   \midrule
    $\diffkze = \mixlen \wvelq$ & \emph{Concentration-profile}: Balance between upward buoyant drift and downward turbulent mixing & Constant $\diffk$; diffusive mixing; constant $\wvelq$ & $\concavg(\zDir)$; $\wvelq$
 \\
 \midrule
    $\diffkzv = \frac{\eddyVisc}{\Sct}$ & \emph{Viscosity}: Particle turbulent diffusivity proportional to momentum turbulent diffusivity & Gradient diffusion holds; known eddy viscosity  & $\eddyVisc(\zDir)$; $\Sct$

  \\
    \bottomrule
  \end{tabularx}
  \label{tab:methods}
\end{table}
The approaches above estimate the same vertical turbulent diffusivity $\diffk$ from different data under different assumptions.
The Lagrangian method ($\diffkzd$) uses particle trajectories and requires no model for the particle drift velocity or turbulent flux.
The Eulerian approaches ($\diffkzm$, $\diffkze$, $\diffkzv$) use  concentration profile data and/or velocity statistics and assume a gradient diffusion model.  
$\diffkzm$ and $\diffkze$ also assume a prescribed drift velocity (typically $\riseVel=\wvelq$). 
Thus, these estimates can fail if the particle's effective rise velocity deviates from its quiescent value, or if the nonlocal diffusion becomes important. 
Table \ref{tab:methods} summarizes the core formulas, underlying physical principles, necessary assumptions, and required input data for each method we test in our study.

\section{Methodology}
\label{sec:Methodology}

\subsection{Experimental setup}
\label{subsec:expset}
Our experiments take place in the Washington Air-Sea Interaction Research Facility, a water channel with a test section 12.2 m long, 0.91 m wide, and 1.22 m high. 
The channel is filled with tap water to a depth of 0.6 m leaving 0.6 m of headspace for airflow from the closed-loop wind tunnel above.
Wind stress on the surface generates a young sea state with deep-water waves and drives a slow mean current.
We install a sloped porous beach at the downstream end to absorb wave energy and to minimize reflections.

Because we are studying particle vertical dispersion, we require observations of vertical particle displacements over long times. Thus,  we need to measure particle trajectories over a sufficiently large field of view. We use large-scale shadow tracking (LSST), a technique introduced in \citet{Baker_2023}, which measures 2D projections of particle trajectories within a 3D volume by tracking particle shadows from a collimated light source. 
With this technique, we can directly measure the particles’ streamwise and vertical positions ($\xDir$ and $\zDir$), while the spanwise ($\yDir$) position remains unknown.  
The LSST system uses four cameras to form a 1~m $\times$ 0.45~m field of view with an average resolution of 9 pixels/mm such that we can resolve the shadows of particles with diameters as small as 2 mm. 
The light source is a 470 nm light-emitting diode (LED) collimated through a large Fresnel lens (1 m $\times$ 0.7 m) positioned on the glass sidewall of the tank. 
The field of view is located 7~m downstream from the test-section inlet, and we capture images at a rate of 30 Hz. We only can measure submerged particles, and therefore trajectories terminate once particles reach the surface. 
For further setup and equipment details, see \citet{Baker_2023}. 

Because LSST only measures projections of particle trajectories, we characterized the flow separately.  
We measured vertical profiles of all three fluid velocity components ($\uVel$, $\vVel$, and $\wVel$) collected at multiple depths using a Nortek Vectrino acoustic Doppler velocimetry  profiler (ADV).
The ADV was mounted above the tank centered within the LSST volume. 
At each measurement point, the ADV resolves the flow at eight vertical cells spaced 4 mm apart, and a Cartesian robot traversed the ADV across the full water depth. At each measurement location, we sampled  at 50 Hz for 2 minutes. 
The ADV's coordinates were rotated into the laboratory frame using a rotation matrix obtained from a singular value decomposition of calibration-run velocity measurements, with the principal axis of maximum variance defining the streamwise direction.
We denoised the velocity signals using signal-to-noise ratio and correlation thresholds of 5 and 40, respectively, and rejected measurements within 6 mm of the free surface (detected by the acoustic backscatter) to avoid contamination by surface echoes.
We additionally measured the free-surface elevation $\wfreesurf$ using a single-point U-GAGE T30X acoustic wave gauge, located at the upstream edge of the LSST volume. 
It recorded continuously for 5 minutes at a sampling rate of 400 Hz, decimated to 100 Hz during postprocessing.

\begin{table}
\centering

\centering
\begin{tabular}{lSS}
\toprule
Wind speed (m/s) & 12 & 16 \\
\midrule
$\waterdepth$ (cm) & 60.0(10) & 60.0(10) \\
$\sigwaveheight$ (cm) & 3.0(9) & 5.3(9) \\
$\wavepeakfreq$ (Hz) & 2.34(2) & 2.05(2)\\
$\wavephsp$ (m/s) & 0.67(2) & 0.76(3)\\
$\peakwavenum$ (rad/m) & 22.04(4) & 16.91(4)\\
$\peakwavelen$ (m) & 0.29(2) & 0.37(2)\\
$\peakwavenum \waterdepth$ (--) & 13.22(50) & 9.21(50)\\
$\velSto$ (cm/s) & 7.5(2) & 15.6(2)\\
\midrule
$\shearVel$ (cm/s) & 2.6(4) & 3.2(4)\\
$\La$ (--) & 0.6(2) & 0.5(2) \\
\bottomrule
\end{tabular}

\caption{Wave and turbulence parameters for experiments conducted at wind speeds of 12 and 16 m/s. 
Wave properties and uncertainties are estimated from the vertical velocity spectra, while turbulence statistics and uncertainties are computed from near-surface fluid velocity time series.  }
\label{tab:summaryflow}

\end{table}

\begin{figure}
    \centering
    \includegraphics[scale=0.4]{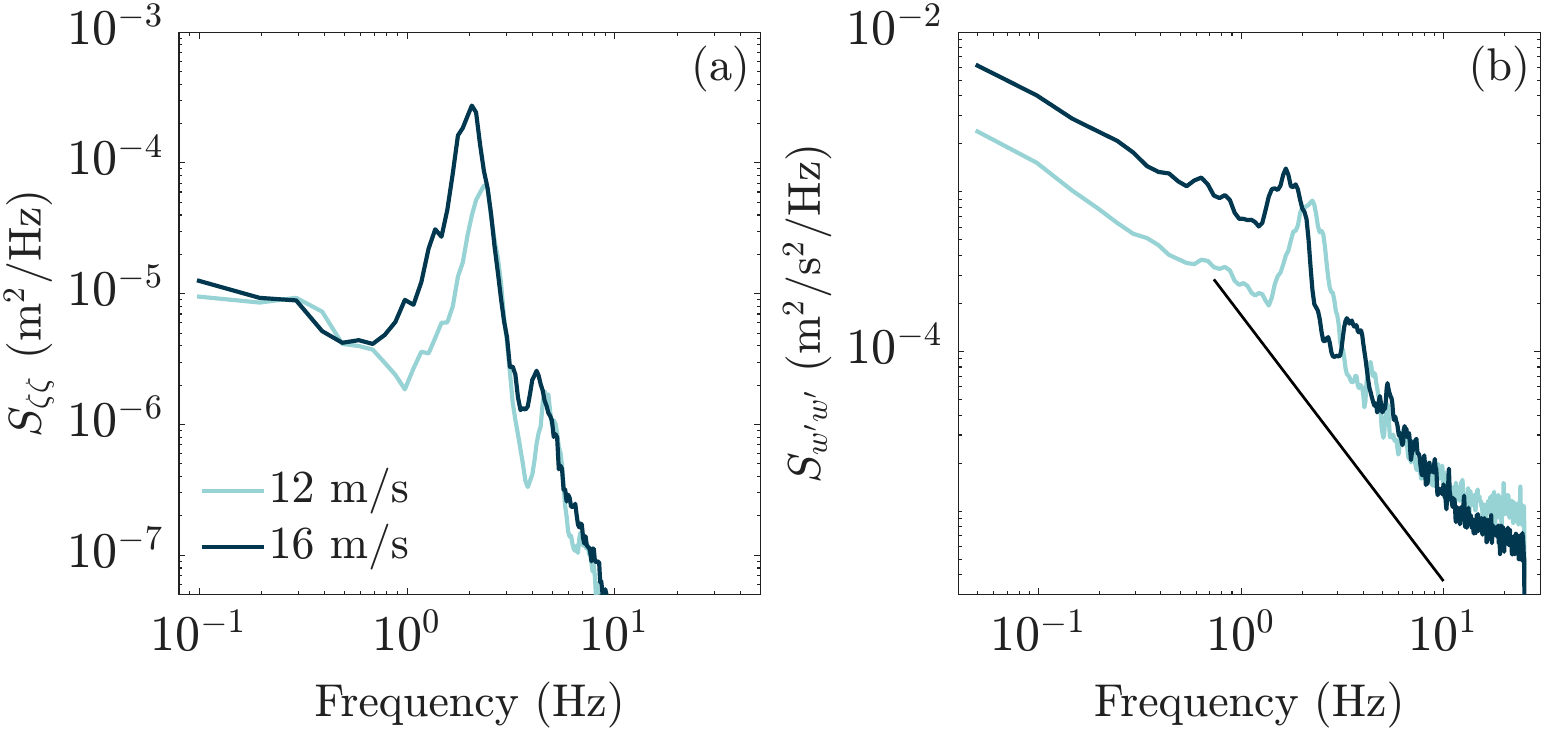}
    \caption{Power spectrum of (a) surface elevation $\waveampspec$ and (b) depth-averaged vertical velocity $\wvelpspec$ for experiments conducted under 12 m/s and 16 m/s wind conditions. The straight black line in (b) has a $-5/3$ slope.}
    \label{fig:surfspectra}
\end{figure}

\subsection{Flow characterization}
\label{subsec:flow}
Here, we characterize both the turbulence and the wave field. 
We plot the power spectral density for the surface elevation $\waveampspec$ and the vertical velocity fluctuations $\wvelpspec$ (depth-averaged over the upper 30 cm) in Figure \ref{fig:surfspectra} for each wind speed.
The spectra were estimated using Welch’s method (segment averaging with a Hanning window and $50\%$ overlap).
The elevation spectra (Figure \ref{fig:surfspectra}a)  peak at 2.34 Hz (12 m/s) and 2.05 Hz (16 m/s). 
The  velocity  spectra (Figure \ref{fig:surfspectra}b)  peak at 2.25 Hz (12 m/s) and 1.66 Hz (16 m/s). 
Note that the velocity spectra peak at slightly lower frequencies because of the depth averaging. 
The 16 m/s wind speed case has a broader spectrum with more energy at lower frequencies, consistent with a more developed wave field.
The higher harmonics are consistent with bound harmonics of the steep waves at the peak frequency, but also potentially representative of wave-modulated turbulence \citep{Thais_1996, Guo_2013}. 
Figure \ref{fig:surfspectra}b also shows a $-5/3$ power-law decay at higher frequencies, characteristic of a developed turbulent inertial range.

The peak frequency $\wavepeakfreq$ is identified as the peak of the surface-elevation power spectrum $\waveampspec(\wavefreq)$ in Figure \ref{fig:surfspectra}a. We also compute the peak angular wave frequency $\jonangfreqpeak=2\pi \wavepeakfreq$, the peak wavenumber $\peakwavenum = \jonangfreqpeak^2 / \gmag$ (using the deep-water wave dispersion relation),  the peak wavelength $\peakwavelen=2 \pi/\peakwavenum$, and the peak wave phase speed $\wavephsp = \gmag/\jonangfreqpeak$. 
We also measure the significant wave height $ \sigwaveheight = 4(\int \waveampspec (\wavefreq) \,\mathrm{d}\wavefreq)^{1/2}$. 
These parameters are reported in the upper part of table \ref{tab:summaryflow}; their uncertainties were obtained from the $95\%$ confidence interval of the averaged spectral estimates.

\begin{figure}
    \centering
    \includegraphics[scale=0.4]{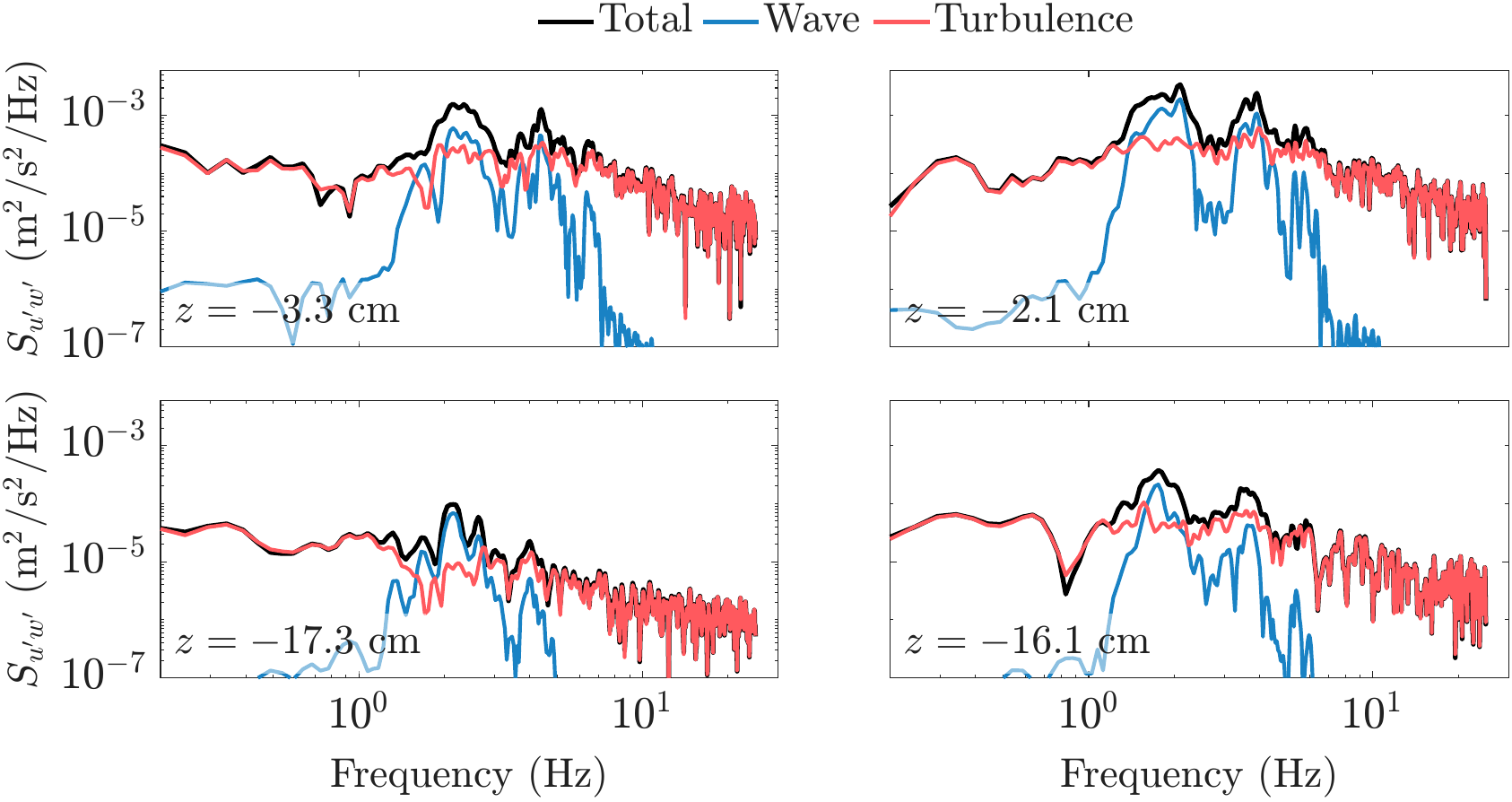}
    \caption{Power spectrum of Reynolds stress from the raw, wave-, and turbulence-only contribution used to assess the effectiveness of the DMD-based separation algorithm. The spectra were checked at two different depths, seen in the lower left corner of each subfigure, for 12 m/s (left column) and 16 m/s (right column).}
    \label{fig:uwdmd}
\end{figure}

Next, we  calculate the mean and turbulent flow statistics. 
We obtain the mean velocity profiles $\uavg$ and $\wavg$ directly from the velocity time series. However, to characterize the turbulent fluctuations, we need to first isolate them from the wave fluctuations. 
Our approach is to remove wave-induced motion from the velocity signals using a dynamic mode decomposition (DMD)-based wave-turbulence separation method that we developed in \citet{Chavez-Dorado_2025}. 
Given a 1-D velocity time series, this separation method first constructs a Hankel-like time-delay matrix to overcome dimensionality limitations of DMD and then applies DMD to decompose the matrix into a set of modes with well‑defined complex frequencies where the coherent modes around the dominant wave frequency are identified as the wave component. 
Subtracting these reconstructed wave modes from the original time-series yields a residual time-series dominated by turbulence.
For our data, we constructed time-delay matrices with shape $4000 \times 2000$ from the $\uVel$ and $\wVel$ time series and reconstructed the wave signal from the first 200 DMD modes. 
Figure \ref{fig:uwdmd} shows that this separation algorithm is able to successfully isolate the wave frequency band including higher harmonics, while preserving the turbulence.
From these turbulence velocity signals we quantify profiles of the horizontal and vertical velocity fluctuation root-mean-square (RMS) $\uturbrms$, $\wturbrms$ and the Reynolds stress $\langle \uf \wf \rangle$.

\begin{figure}
    \centering
    \includegraphics[scale=0.41]{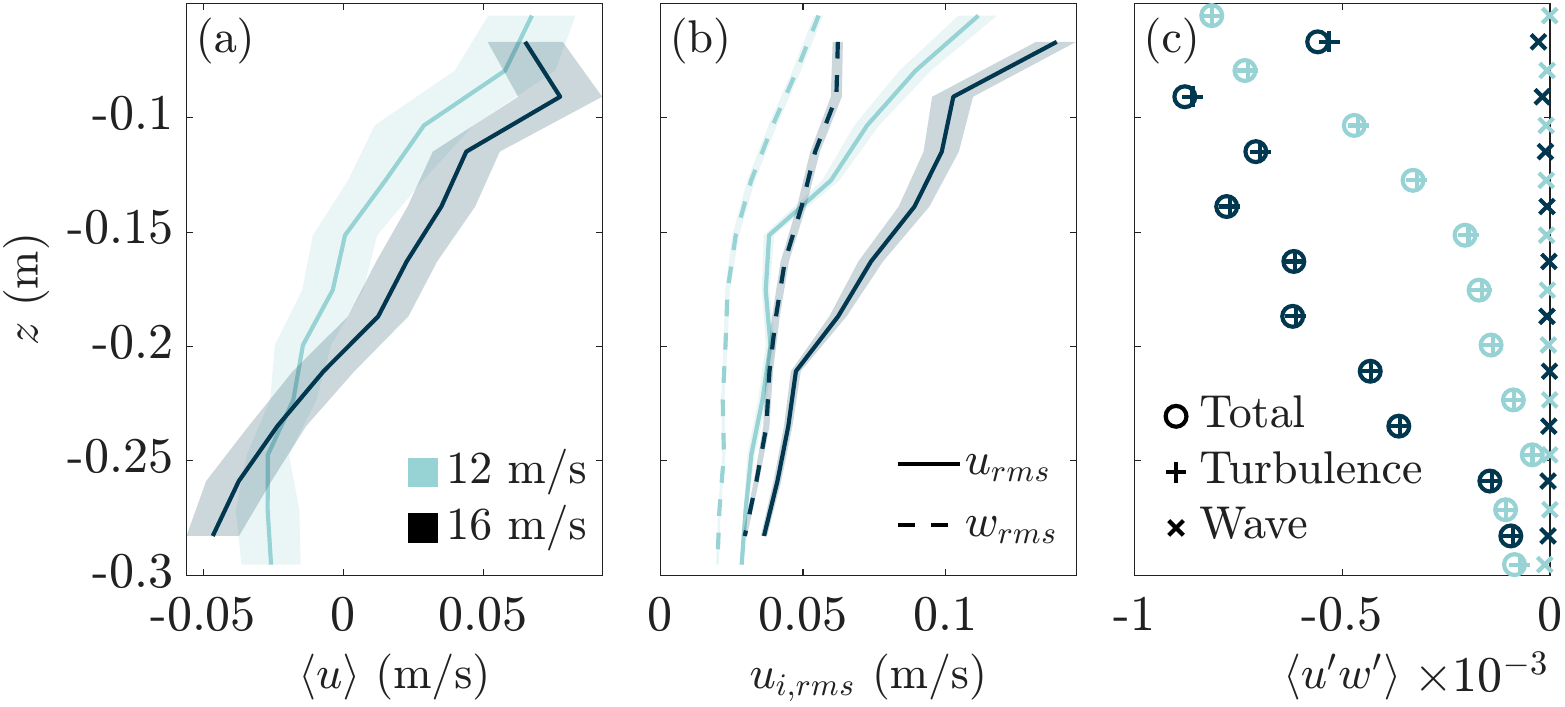}
    \caption{(a) Time-averaged streamwise and vertical velocity profiles $\uavg(\zDir)$ and $\wavg(\zDir)$, respectively. (b) Streamwise and vertical velocity turbulent RMS profiles $\uturbrms (\zDir)$ and $\wturbrms(\zDir)$, respectively. (c) Reynolds stress profiles showing the waves and turbulence contributions separated using the DMD separation algorithm. The shaded regions in all figures show the $95\%$ confidence interval derived from bootstrapping. }
    \label{fig:meanVel_PIV2023}
\end{figure}

We plot the mean horizontal velocity profile in Figure \ref{fig:meanVel_PIV2023}a where we see a sheared mean current with a reverse current near the bottom that sets up due to the confinement of the tank. 
For this reason, we restrict most of our analysis to the upper 30 cm of the water column.
Below this depth, the mean velocity gradient (needed for the eddy viscosity estimation) approaches zero and changes sign because of the flow reversal, making the eddy viscosity physically uninterpretable.
We apply the same depth restriction to all statistics to maintain a consistent sampling domain.
In Figure \ref{fig:meanVel_PIV2023}b we plot the vertical profiles of turbulent RMS velocities, showing the highest turbulent fluctuations near the surface. 
We plot the Reynolds shear stresses over depth in Figure \ref{fig:meanVel_PIV2023}c, showing both the components due to the waves and the turbulent fluctuations separately. As expected, the main contributions to the shear stress come from the turbulence and not the waves, consistent with similar wind-driven flows \citep{Cheung_1988, Thais_1996, Terray_1996}. Note that this result also helps illustrate the validity of our wave-turbulence decomposition. 

\begin{figure}
    \centering
    \includegraphics[scale=0.4]{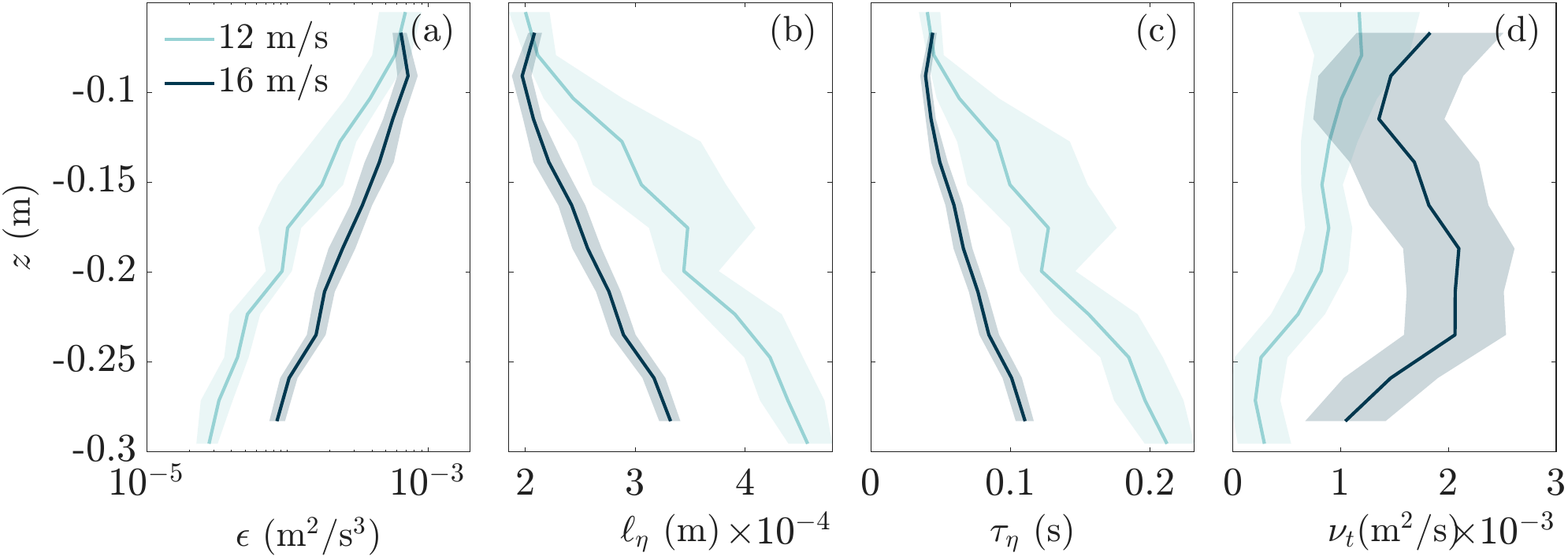}
    \caption{Vertical profiles of (a) turbulent dissipation using \citep{Gerbi_2009}, (b) Kolmogorov length scale and (c) time scale and (d) eddy viscosity from \eqref{eqn:eddyviscgrad}.}
    \label{fig:dissipgerbi}
\end{figure}

In Figure \ref{fig:dissipgerbi} we plot estimates of the turbulent dissipation, Kolmogorov length- and time-scales, and eddy viscosity. We estimate the dissipation rate $\dissturb$ as a function of depth using inertial-subrange spectral fitting following \citet{Terray_1996} and \citet{Gerbi_2009}. 
Given the kinematic viscosity $\visck = 10^{-6}$ m$^2$/s, we also calculate the Kolmogorov lengthscale $\kls =\left( \frac{\visck^3}{\dissturb} \right)^{1/4}$ and timescale $  \rtimef =\left( \frac{\visck }{\dissturb} \right)^{1/2}.$ We estimate eddy viscosity $\eddyVisc$ with equation \ref{eqn:eddyviscgrad}.
As expected, we see that in Figure \ref{fig:dissipgerbi}, $\dissturb$ decreases with depth given that turbulence is generated near the water surface. Thus, $\kls$ and  $\rtimef$ both increase with depth, indicating that the smallest scale eddies occur near the surface, as expected.
These dissipation values are consistent with field measurements in 14 m/s winds \citep{Gemmrich_2004}. In contrast, we see that  $\eddyVisc$ is relatively constant over depth. 
This differs from a classic wall scaling, but is instead consistent with the more uniform mixing expected under a wave layer \citep[as discussed in ][]{Gerbi_2009, Kukulka_2015, Craig_1994}.

Finally, we characterize the relative importance of waves and turbulence with the Langmuir number $\La = (\shearVel / \velSto)^{1/2}$, defined with the water friction velocity $\shearVel$ and the surface Stokes drift velocity $\velSto$. We estimate the friction velocity $\shearVel = \sqrt{|\langle \uf \wf \rangle|}$ 
using values measured at 5~cm below the mean water level to avoid contamination by the waves.  
The Stokes drift velocity is estimated from the wave amplitude spectrum: 
\begin{equation}
    \velSto = 2 \int \jonangfreq \wavenum \waveampspec(\wavefreq) \,\mathrm{d}\wavefreq,
\end{equation}
as in \citet{Kumar_2017}. Small values of $\La$ indicate that waves dominate over shear-driven turbulence, whereas large values indicate a shear-dominated regime.
For reference, we report $\shearVel$, $\velSto$, and $La$ for each wind speed  in table \ref{tab:summaryflow}. 
We find $\La$ values of $0.5-0.6$, indicating wave influence, but not necessarily fully-developed Langmuir circulation. For reference, \cite{Li_2005} concludes that, for a developed flow, Langmuir turbulence dominates for $\La \lesssim 0.3$.
Compared to equilibrium open-ocean conditions which often have more wave influence with $\La \approx 0.2 \text{--}0.5$ \citep{Fisher_2024}, our values are near the shear-dominated regime. 
In our experiments, the $\La$ values suggests that Langmuir turbulence is not expected to be a strong effect.

\subsection{Particles}
\label{subsec:particles}

\begin{sidewaystable}

\begin{tabular}{l l *{9}{S} }
\toprule
Wind & Parameter
& {Small} & {Medium} & {Large}
& {Small} & {Medium} & {Large}
& {Medium} & {Large}
& {Neutral} \\
speed &
& {spheres} & {spheres} & {spheres}
& {disks} & {disks} & {disks}
& {rods} & {rods}
& {} \\
\midrule
\multirow{6}{*}{0 m/s}
& $\partDiam$ (mm) & \num{2.0(0.1)} & \num{2.5(0.1)} & \num{3.9(0.3)} & \num{3.1(0.1)} & \num{3.9(0.1)} & \num{4.9(0.1)} & \num{3.7(0.1)} & \num{4.6(0.2)} & \num{2.7(0.3)} \\
& $\wvelq$ (cm/s) & \num{2.7(1)} & \num{3.7(1)} & \num{5.5(1)} & \num{2.3(1)} & \num{2.5(1)} & \num{2.6(1)} & \num{3.5(1)} & \num{3.7(1)} & \num{0.1(1)}\\
& $\denp$ (kg/m$^3$) & \num{960(1)} & \num{960(1)} & \num{960(1)} & \num{960(1)} & \num{960(1)} & \num{960(1)} & \num{960(1)} & \num{960(1)} & \num{997(1)}\\
& $\rtimep$ (ms) & \num{97(3)} & \num{136(4)} & \num{203(6)} & \num{86(3)} & \num{92(3)} & \num{97(5)} & \num{129(4)} & \num{136(4)} & \num{69(4)}\\
& $\Rep$ (--) & \num{60(3)} & \num{104(6)} & \num{243(27)} & \num{71(7)} & \num{95(9)} & \num{301(10)} & \num{129(5)} & \num{128(7)} & \num{4(3)}\\
& $\Ga$ (--) & \num{69(5)} & \num{97(5)} & \num{190(17)} & \num{114(5)} & \num{151(6)} & \num{216(7)} & \num{134(3)} & \num{217(3)} & \num{22(4)} \\
& $\denpar$ (--) & \num{1.02(1)} & \num{1.02(1)} & \num{1.02(1)} & \num{1.02(1)} & \num{1.01(1)} & \num{1.01(1)} & \num{1.02(1)} & \num{1.02(1)} & \num{1.00(1)}\\

\midrule

\multirow{3}{*}{12 m/s}
& $\St_c$ (--) & \num{0.3(1)} & \num{0.4(1)} & \num{0.6(1)} & \num{0.3(2)} & \num{0.2(2)} & \num{0.2(2)} & \num{0.4(1)} & \num{0.4(1)} & \num{0.1(1)} \\

& $\Sv$ (--)
& \num{0.9(3)} & \num{1.2(3)} & \num{1.9(5)} & \num{0.7(1)} & \num{0.8(3)} & \num{0.8(3)} & \num{0.9(5)} & \num{1.1(4)} & \num{0.06(30)} \\
& $\Sz$ (--)
& \num{9(3)} & \num{11(3)} & \num{22(3)} & \num{13(4)} & \num{15(5)} & \num{19(6)} & \num{16(4)} & \num{20(6)} & \num{9(3)} \\

\midrule

\multirow{3}{*}{16 m/s}
& $\St_c$ (--) & \num{0.4(1)} & \num{0.4(2)} & \num{0.5(1)} & \num{0.3(1)} & \num{0.2(1)} & \num{0.2(1)} & \num{0.3(1)} & \num{0.3(1)} & \num{0.2(1)} \\

& $\Sv$ (--)
& \num{0.6(2)} & \num{0.8(2)} & \num{1.0(3)} & \num{0.5(2)} & \num{0.5(1)} & \num{0.5(2)} & \num{0.7(2)} & \num{0.7(3)} & \num{0.03(10)} \\
& $\Sz$ (--)
& \num{9(2)} & \num{12(2)} & \num{20(3)} & \num{13(3)} & \num{16(3)} & \num{21(4)} & \num{16(3)} & \num{20(3)} & \num{12(3)} \\

\bottomrule
\end{tabular}

\caption{Summary of the particle characteristics in quiescent water and wind-driven surface boundary layer for 12 and 16 m/s.
For quiescent conditions (0 m/s), the error bounds represent the measurement tolerances for particle physical properties (such as diameter and density) and confidence intervals of terminal rise velocity across multiple releases of particles in still water. The uncertainty for the non-dimensional parameters is calculated through error propagation.
For wind-driven conditions ($>0$ m/s), the error bounds represent standard deviation due to the spread of the particles' vertical positions through the water column.}

\label{tab:summaryparticlesflow}
\end{sidewaystable}

We use buoyant particles chosen to target an intermediate regime where their buoyancy is strong enough relative to the turbulence that they do not behave as uniformly-mixed tracers, yet weak enough that the turbulence is able to mix them below the free surface.  
The particles are high-density polyethylene (HDPE) particles  with specific gravity $\SG = \denp/\denf= 0.96$ in three shapes (spheres, rods and disks). We also use particles that we made out of Carnauba‑wax ($\SG = 0.997$) that approximate a tracer and hereafter are referred to as \emph{neutral} particles.

The particles' shape is characterized by the aspect ratio $\axar = \arsym/\arper$, where $\arsym$ denotes the length of the axis of rotational symmetry and $\arper$ the characteristic length of the perpendicular axes. 
The spheres (aspect ratio $\axar=1$) have diameters of 2.0 mm, 2.5 mm (precision spheres, Cospheric) and 3.9 mm (nurdles, McMaster‑Carr).
The rods (prolate, $\axar>1$) were made by cutting 3D‑printer filament with a 1.75 mm diameter into segments with average length 10.7 mm and 20.2 mm, giving aspect ratios of 6.1 and 11.5. 
The disks (oblate, $\axar<1$) were die-cut from 0.79 mm-thick HDPE sheets  with diameters of 5.0 mm, 7.0 mm and 10.0 mm, giving aspect ratios of 0.16, 0.11 and 0.08. 
The neutral particles are cylinders with both an equal height and diameter of 2.7 mm ($\axar=1$).
We additionally define the nonspherical particles by their volume‑equivalent diameter, $\partDiam = (1.5\arper^{2}\arsym)^{1/3}$. 
The particle properties are summarized in table \ref{tab:summaryparticlesflow}.
Because the particles are hydrophobic, before each experiment we soak them in water with a few drops of dish soap to prevent air bubbles from adhering when we introduce them to the tank.

We measured the particles' terminal rise velocity in quiescent water.
We released particles one at a time in a rectangular tank measuring 25 cm $\times$ 25 cm $\times$ 100 cm high. 
For each particle shape and size, 15 realizations were performed by selecting particles at random with replacement from a set of 20 (where each particle class had at least 50 total particles). 
We lowered the individual particles to the bottom of the tank using a claw grabber and released it after the fluid had come to rest. 
The rods and disks oscillated laterally during their rise due to shedding vortices given their intermediate Reynolds number. 
We recorded the particles at 30~fps over a 15 x 15 cm field of view centered 15 cm below the free surface. 
We computed the time-averaged vertical velocity of each realization within the field of view, checking that the temporal standard deviation was small relative to the mean (less than $10\%$ for all cases) ensuring that the particles had reached their terminal velocity.
We report the average terminal velocities of the particles in Table \ref{tab:summaryparticlesflow}. 

We characterize the particles and estimate relevant non-dimensional parameters in both still fluid and the turbulent flow in the experiments. We begin with the still fluid characterization, calculating both  the particle Reynolds number $\Rep=\wvelq\partDiam/\visck$ and the Galileo number $\Ga=\sqrt{|\denp /\denf-1|\,\gmag\,\partDiam^3}/\visck$. 
$\Rep$ quantifies the relative importance of inertial to viscous forces in the fluid surrounding a particle, and $\Ga$ quantifies the relative importance of particle buoyancy to viscous forces.
We also compute the density parameter $\denpar = (1 + \amasscoeff) / (\denp/\denf + \amasscoeff)$, where $\amasscoeff$ is the orientation-averaged added mass coefficient.
For each particle shape, we estimate the $\amasscoeff$ from the classical solution for spheroids in potential flow \citep{Lamb_1930}.
We report these values for all particle geometries and wind speeds in table \ref{tab:summaryparticlesflow}.
The $\Rep$ ranges from 4 (neutral) to 301 (large disk) and indicates that for many of the cases, the drag is in a nonlinear regime. This range is similar to reported values for real microplastics, which also have variable shape \citep{DiBenedetto_2023}. 
For the nonspherical particles, the Galileo numbers ($\Ga = 69\text{--}217$) indicates path instabilities \citep{Fernandes_2007}, consistent with the wobbling trajectories observed during the terminal velocity measurements.
The resulting $\denpar$ values fall within a narrow range close to one, which means that particle inertia effects in turbulence are likely to be minor, as shown in \citet{Berk_2024}.

Next, in order to compare each particle with the turbulent flow statistics it actually experiences, we define  effective flow scales sampled by the particles. We weight the vertical profiles of $\wturbrms$, $\kls$, and $\rtimef$ (section \ref{subsec:flow}) by the measured particle vertical distribution  $\wturbrmse=\langle \wturbrms(\zplagr(\tm))\rangle,$ $\klse=\langle \kls(\zplagr(\tm))\rangle,$  {and}  $\rtimefe=\langle \rtimef(\zplagr(\tm))\rangle$, accounting for the particles' non-uniform  distribution with depth.
These are effective mean scales sampled by the particle ensembles rather than instantaneous values, and thus will serve as the basis for comparing bulk transport statistics across particle types.

From these effective flow scales, we define three non-dimensional parameters that govern the particle-turbulence interactions: the non-dimensional rise velocity  $\Sv = \wvelq/\wturbrmse$ (similar to a Rouse number), the relative particle size $\Sz = \partDiam/\klse$, and the Stokes number $\St = \rtimep/\rtimefe$. 
Here, we defined the particle relaxation time with the measured terminal rise velocity: $\rtimep = \wvelq/((1 - \denpar)g)$.
We report the range of values in table \ref{tab:summaryparticlesflow}, where the uncertainty is a measure of the particle sampling spread along the water column.
We find that particle buoyancy is comparable to or stronger than turbulent fluctuations ($\Sv = 0.5 \text{--} 1.9$ for buoyant particles), whereas the neutral tracer has $\Sv\approx 0$. 
Also, the particles are much larger than the Kolmogorov length ($\Sz \approx 9 \text{--} 22$), so finite‐size effects may be important for all shapes.
Given the finite size, the Stokes number based on the Kolmogorov time scale $\rtimefe$ is not an appropriate measure of particle inertia since particles with diameter $\partDiam$ are more strongly coupled with eddies of comparable size \citep{Xu_2008}.
Following \citet{Xu_2008}, we define a corrected Stokes number $\St_c = \St/\Sz^{2/3}$, which replaces the Kolmogorov time scale by the eddy turnover time at the particle scale $(\partDiam^2/\dissturb)^{1/3}=\rtimefe\Sz^{2/3}$.
The corrected Stokes numbers range from $0.1$ to $0.6$, showing that particle inertia is not strong relative to the eddies at the particle scale.

\subsection{Particle experiments}

Before each run, we let the wind run for approximately 15 minutes to ensure that the flow reached steady state.
Each run lasted 2 minutes, and we released approximately 70-90 particles per run for each particle type.
We released particles immediately below the water surface at  $\Delta x\approx 5$ m upstream of the field of view.
Before the particles enter the field of view, they must also have enough time to reach their vertical equilibrium state.
To verify this, we compare the horizontal advective transit time $T_{adv} = \Delta x / \hat{\langle u \rangle}$  with the turbulent mixing time scale $T_{mix} = \mixlen^2/\diffkzd$. 
Here, $\hat{\langle u \rangle}$ is the streamwise velocity.
If $T_{mix} < T_{adv}$, the particles have enough time to reach an equilibrium state before being measured.
Using the 16 m/s wind speed case as a conservative estimate, we have $\hat{\langle u \rangle} \approx 5$ cm/s near the surface, giving $T_{adv} \approx 100$ s.
For $T_{mix}$, we use the particles with the largest $\mixlen^2/\diffkzd$ because they are the slowest to reach equilibrium.
From section \ref{sec:Results}, we have $T_{mix} \approx 77$ s.
Thus, even the slowest particles should have enough time to reach equilibrium by the time they reach the experimental field of view.

Beyond this scaling analysis, and given that our analysis and estimates of diffusivity from the particle concentration profiles rely on assumptions of horizontal homogeneity and steady state, we compare the concentration profiles upstream and downstream in the field of view and at different times in the experiment to confirm these assumptions.
For the few instances where homogeneity or steady-state conditions were not met for certain particle types, the non-steady portions of the data have been omitted from the analysis.
For a detailed discussion see Appendix \ref{app:steady}.

\section{Results \& Discussion}
\label{sec:Results}
In this section, we present our results for vertical diffusivity of buoyant particles using (1) a Lagrangian approach via the asymptotic estimation of diffusivity $\diffkzd$ obtained from the particle MSD; (2) an Eulerian approach via two estimations from the  particle concentration profiles:  $\diffkze$ from an assumed mean profile and $\diffkzm$ from the mean flux profile; and (3) the estimation $\diffkzv$ from the eddy viscosity $\eddyVisc$  which also serves as a reference for estimating the Schmidt number.

\subsection{Diffusivity measured asymptotically from particle dispersion}
\label{subsec:Asymptotic}
We first estimate the turbulent diffusivity directly from the time evolution of the particles' vertical dispersion.
This Lagrangian estimate is a direct, kinematic measurement and requires no assumptions about either the particle rise velocity or the turbulent flux. 
This will serve as the reference diffusivity against which we test the Eulerian, concentration profile-based methods in \S\ref{subsec:results_conc}.

\begin{figure}
    \centering
    \includegraphics[scale=0.43]{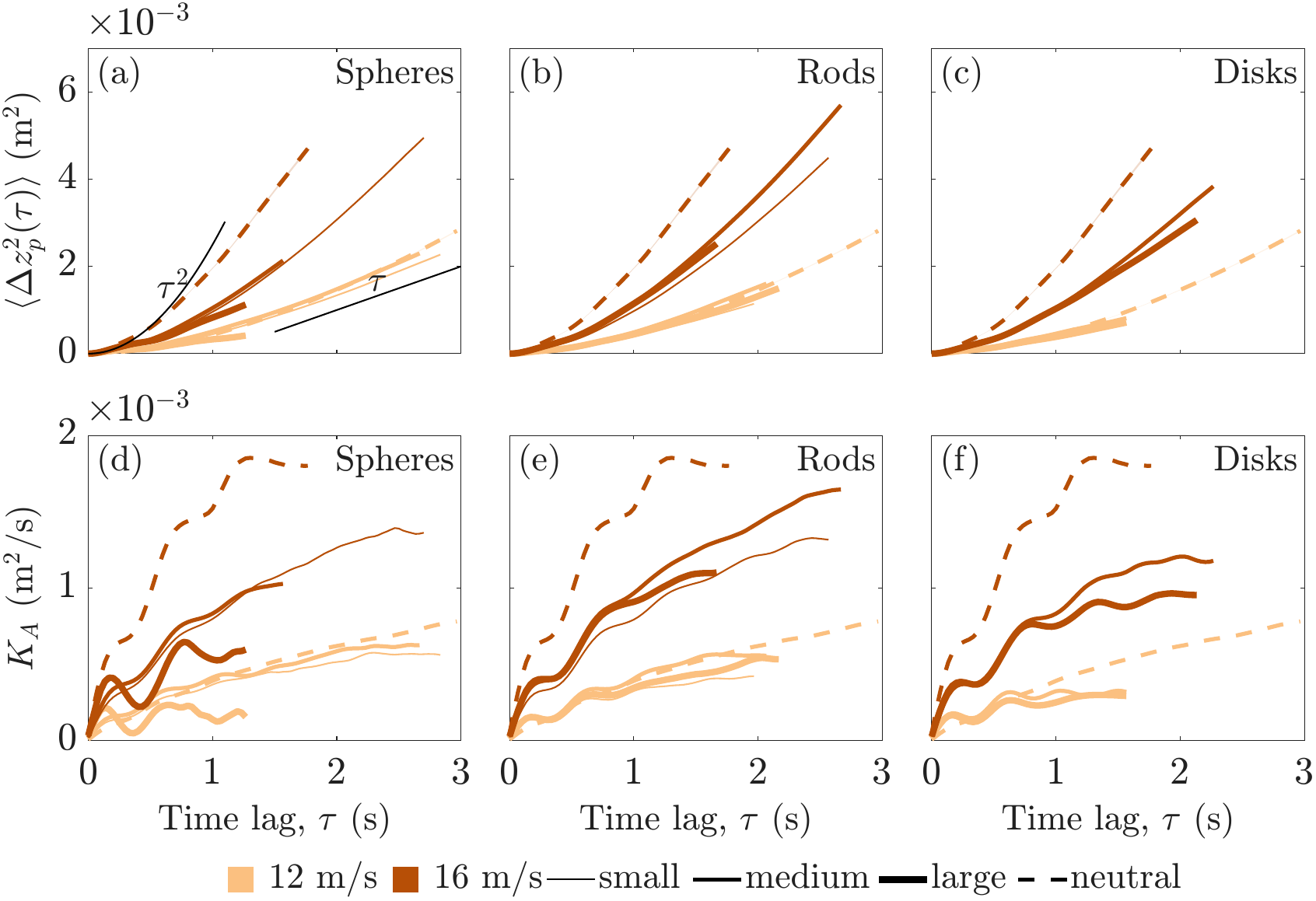}
    \caption{Time evolution of (a-c) particle vertical dispersion and (d-f) instantaneous diffusivity $\diffkzd = 0.5\, \mathrm{d}\vertdispvar / \mathrm{d}\tlag$ for all experiments.}
    \label{fig:direct_sigma_dsigdt}
\end{figure}

We recall that, as discussed in \S\ref{sec:inertia_crosstraj}, a nonzero ensemble-mean vertical velocity would add a ballistic contribution to the MSD and affect our ability to observe a diffusive regime. 
However, we verified that the particle concentration profiles are statistically steady (see Appendix~\ref{app:steady}). Consistent with this, we measure $\langle w_p \rangle \approx 0$ for all particle types and wind speeds. 
A vanishing mean vertical velocity does not mean that the particles are not rising relative to the fluid. 
In steady state, the upward buoyant drift is exactly balanced by the downward turbulent flux, such that the mean velocity of particles at any depth is zero. 
Thus, the Taylor method described in \S\ref{subsec:Lagrangian_methds} can be applied to measure the diffusivity from the dispersion directly.

To calculate dispersion, each particle trajectory is re‑referenced to its own initial time $\tm_0$ and depth $\zplagr(\tm_0)$, and the vertical displacement at time lag $\tlag$ is defined as $\Delta \zplagr(\tlag) = \zplagr(\tm_0 + \tlag) - \zplagr(\tm_0)$. 
The dispersion $\vertdispvar$ is then obtained by ensemble averaging over all available trajectories. 
Because particle trajectories end when they reach the surface,  longer trajectories are biased toward deeper particles. 
To correct this bias, we implement a weighting scheme where each trajectory's contribution is adjusted at every time lag $\tlag$ such that the sampled distribution matches the  stationary distribution  for each experiment (Appendix~\ref{app:biascorr}). 
The result is an effective, depth-averaged dispersion, rather than a local dispersion at a specific depth.

We plot the vertical dispersion for all particles for both wind speeds in Figure~\ref{fig:direct_sigma_dsigdt}a-c. 
In all cases we observe the dispersion following the expected ballistic to diffusive transition, where at short lags the dispersion scales as $\vertdispvar \propto \tlag^2$ and at long lags the dispersion scales as $\vertdispvar \sim \tlag$, with higher wind speeds corresponding to higher dispersion.
The instantaneous diffusivity $\diffk (\tlag) = {1}/{2}\,{\mathrm{d}\vertdispvar}/{\mathrm{d} \tlag}$ is plotted in Figure~\ref{fig:direct_sigma_dsigdt}d-f.
We see the initial rise in $\diffk$ in the ballistic regime followed by a leveling off in the diffusive regime. 
There are also oscillations in the diffusivity due to the waves which are strongest at short time lags.
Note, the most buoyant particles (the large spheres) have the shortest trajectories  because they spend less time underwater.

\begin{figure}
    \centering
    \includegraphics[scale=0.43]{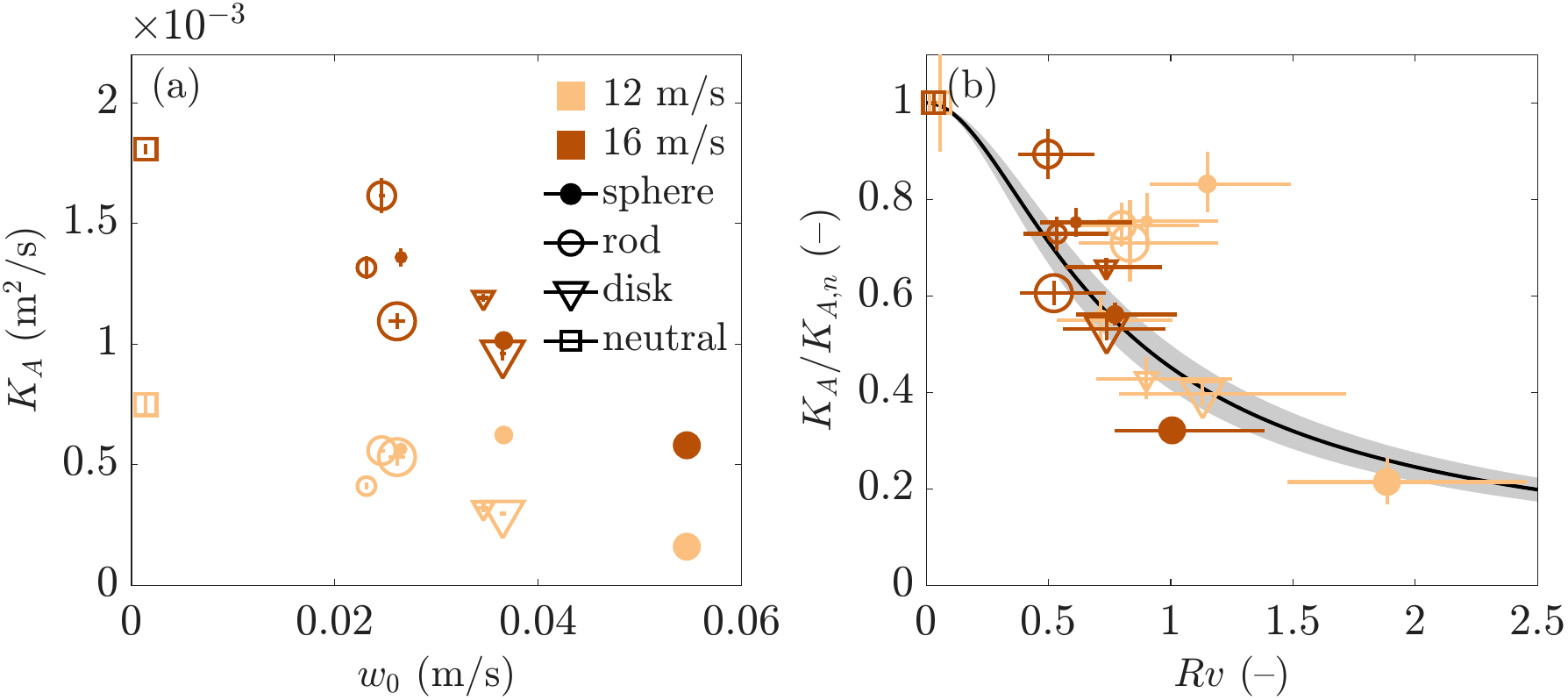}
    \caption{(a) Depth-averaged estimations of long-time diffusivity $\diffkzd$ from the asymptotic method. (b) Ratio of bulk diffusivity from the asymptotic method for each particle class to that of the nearly neutrally buoyant particles, $\diffkzd/K_{A,n}$, versus the velocity parameter $\Sv$. The solid line is Csanady’s prediction in equation \ref{eqn:crosstraj} with $\ctrajfac = 1.9$, where the shaded region represents the 95$\%$ confidence interval of the nonlinear least-squares fit to the data.}
    \label{fig:Kzd_wq}
\end{figure}

We estimate the asymptotic diffusivity $\diffkzd$ by averaging the diffusive plateau of $\diffk(\tm)$ in Figures~\ref{fig:direct_sigma_dsigdt}d-f.
We plot $\diffkzd$ as a function of the particle quiescent rise velocity $\wvelq$ in Figure~\ref{fig:Kzd_wq}a, where 
we see that $\diffkzd$ decreases by a factor of three to four between the neutral particles and the most buoyant ones.
This decrease in diffusivity shows that the more buoyant particles disperse more slowly.
To compare across the different wind speeds, we normalize $\diffkzd$ by the neutral particle value and plot it as a function of $\Sv$ in Figure~\ref{fig:Kzd_wq}b.
Here, the normalized diffusivity $\diffkzd/K_{A,n}$ decreases with increasing buoyancy, and the relationship is well-described by the crossing-trajectories prediction of \citet{Csanady_1963} (Equation \ref{eqn:crosstraj} with best-fit $\ctrajfac=1.9 \pm 0.2$). 
This fitted constant lies within the range reported for slightly buoyant particles in a jet by \citet{Aiyer_2022}, who obtained $\ctrajfac=0.85\text{--}2$. 
Although there is some scatter in the plot, the agreement with the crossing-trajectories theory suggests that  particle rise velocity is the primary control on vertical diffusivity, with particle shape playing only a secondary role and mainly contributing through its effect on $\wvelq$.

\begin{figure}
    \centering
    \includegraphics[scale=0.43]{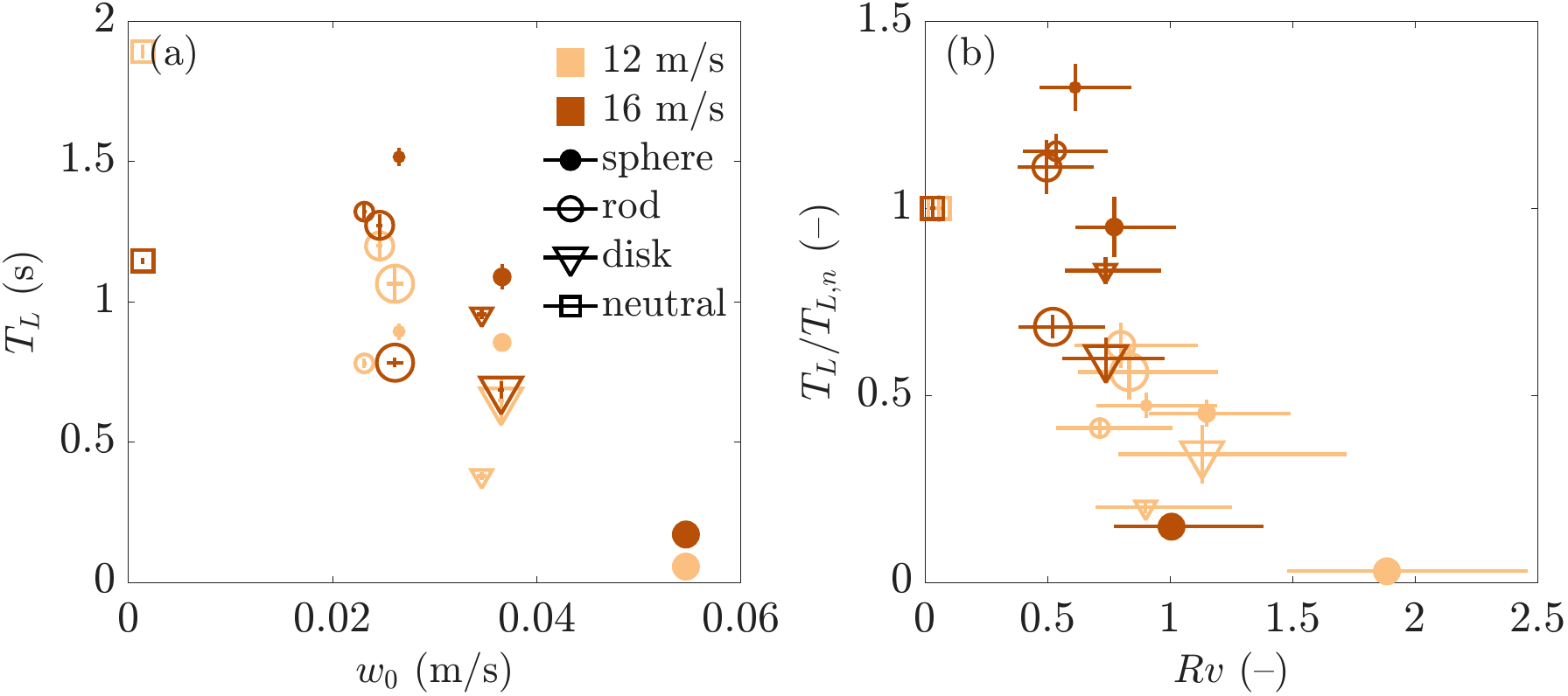}
    \caption{(a) Depth-averaged decorrelation time scale $\lagtsc$ as a function of the quiescent particle rise velocity $w_0$. (b) Decorrelation time scale normalized by the corresponding neutral-particle value, $\lagtsc/T_{L,n}$, as a function of the velocity parameter $\Sv$. Error bars denote the uncertainty of the fitted diffusivity.}
    \label{fig:Tla_wq}
\end{figure}

To further support this hypothesis, we examine the fitted Lagrangian decorrelation time scale $\lagtsc$ by fitting the \citet{Taylor_1922} diffusivity solution
\begin{equation}
    \diffk(\tm)=\frac{1}{2} \frac{\mathrm{d}\vertdispvar}{\mathrm{d}\tlag} =\diffkzd \left(1-e^{-\tlag/\lagtsc} \right),
\end{equation}
to the $\diffkzd$ curves in Figures \ref{fig:direct_sigma_dsigdt}d-f where $\diffkzd$ and $\lagtsc$ are both fitting parameters. 
These fits yield $\diffkzd$ values in strong agreement with the values found from averaging the plateau (Figure \ref{fig:Kzd_wq}a), and therefore we plot only the fitted $\lagtsc$ in Figure \ref{fig:Tla_wq}a as a function of $\wvelq$. 
Here, we see a decrease in $\lagtsc$ with increasing particle rise velocity, confirming that the particles' velocity signal decorrelates increasingly rapidly  as particle drift increases.
Similar to Figure \ref{fig:Kzd_wq}b, we plot the normalized $\lagtsc$ in Figure \ref{fig:Tla_wq}b where we see that it also tends to decrease with $\Sv$. 
Again, this supports our result that the crossing-trajectories effect is responsible for the observed reduction in diffusivity with increasing $\Sv$. 

\subsection{Diffusivity estimated from particle concentration profiles} \label{subsec:results_conc}

\begin{figure}
    \centering
    \includegraphics[scale=0.4]{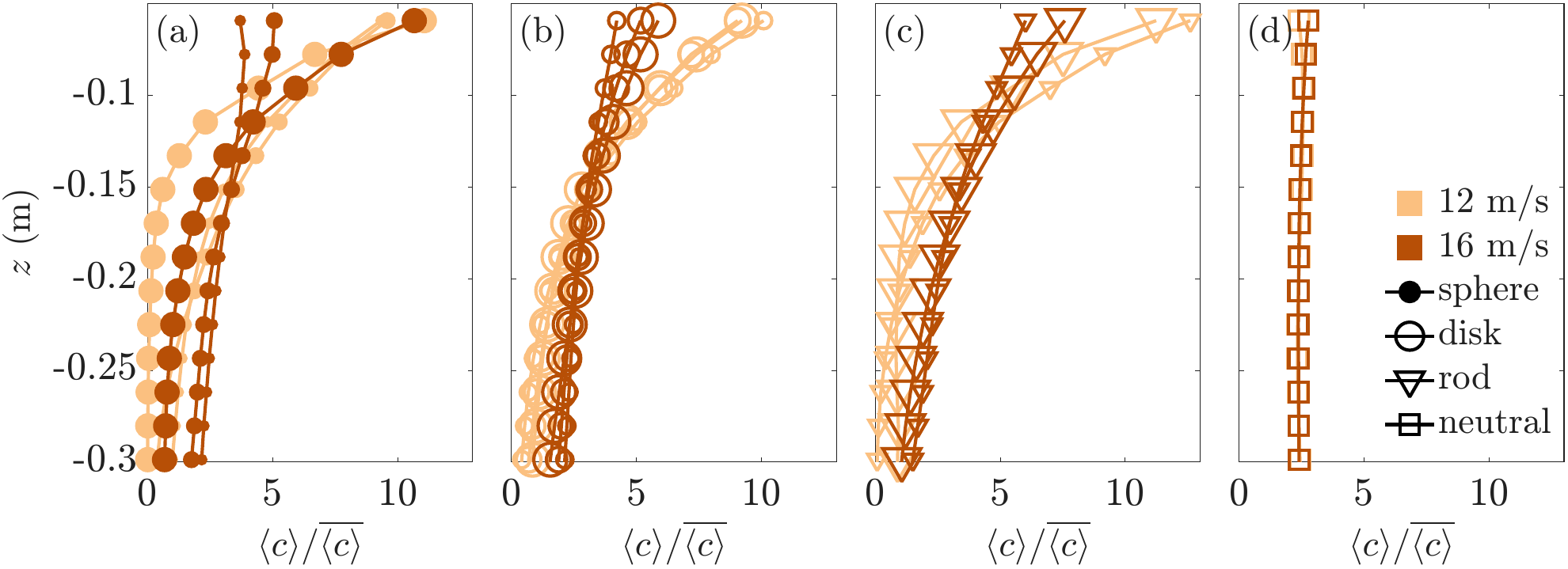}
    \caption{Normalized concentration profiles across different particle types and flow conditions. Each subfigure corresponds to a different particle group. Larger marker sizes indicate larger particle size. 
    }
    \label{fig:conc_prof_normint}
\end{figure}

\begin{figure}
    \centering
    \includegraphics[scale=0.43]{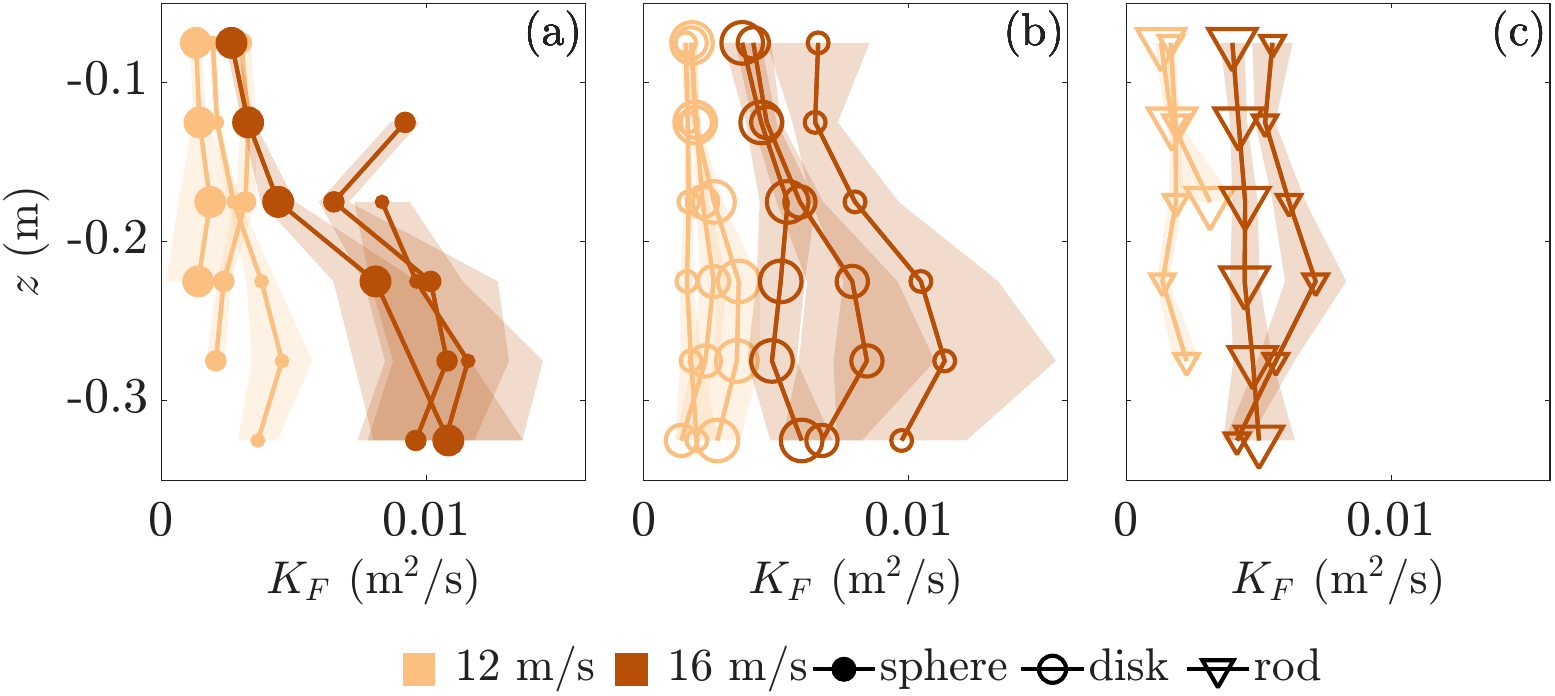}
    \caption{Vertical profiles of Eulerian turbulent diffusivity calculated assuming that the particle effective rise velocity is equal to the quiescent rise velocities. The 95$\%$ confidence intervals were obtained by bootstrap resampling, with uncertainty propagated through the concentration profile and gradient, and diffusivity calculations. We removed measurements with confidence intervals greater than 0.005 m$^2$/s.}
    \label{fig:flux_separate}
\end{figure}

We next evaluate the vertical diffusivity using the Eulerian  concentration profiles $\concavg$ which we have plotted in Figure~\ref{fig:conc_prof_normint} and normalized by the depth-integrated concentration $\overline{\concavg} = \int^0_{-h} \concavg \, d\zDir$. 
In these analyses, we assume the average particle rise velocity to be the terminal velocity in quiescent water $\wvelq$, a common assumption in both sediment and microplastics transport. 
We compare these diffusivity estimates to the more direct Lagrangian estimate $\diffkzd$. 
The neutral particles are excluded from this analysis because they are nearly uniformly mixed; i.e., $\mathrm{d} \concavg/\mathrm{d} \zDir\approx 0$ which leads to large uncertainty in any diffusivity estimate.

We first plot the depth-resolved, flux-estimate diffusivity $\diffkzm$ as defined in \eqref{eqn:diff_f} in Figure~\ref{fig:flux_separate}. 
In all cases, $\diffkzm$ is relatively uniform over depth, especially near the surface where most of the particles are found. 
This supports the assumption of a constant diffusivity over depth, as will be used in the following analysis.

\begin{figure}
    \centering
    \includegraphics[scale=0.43]{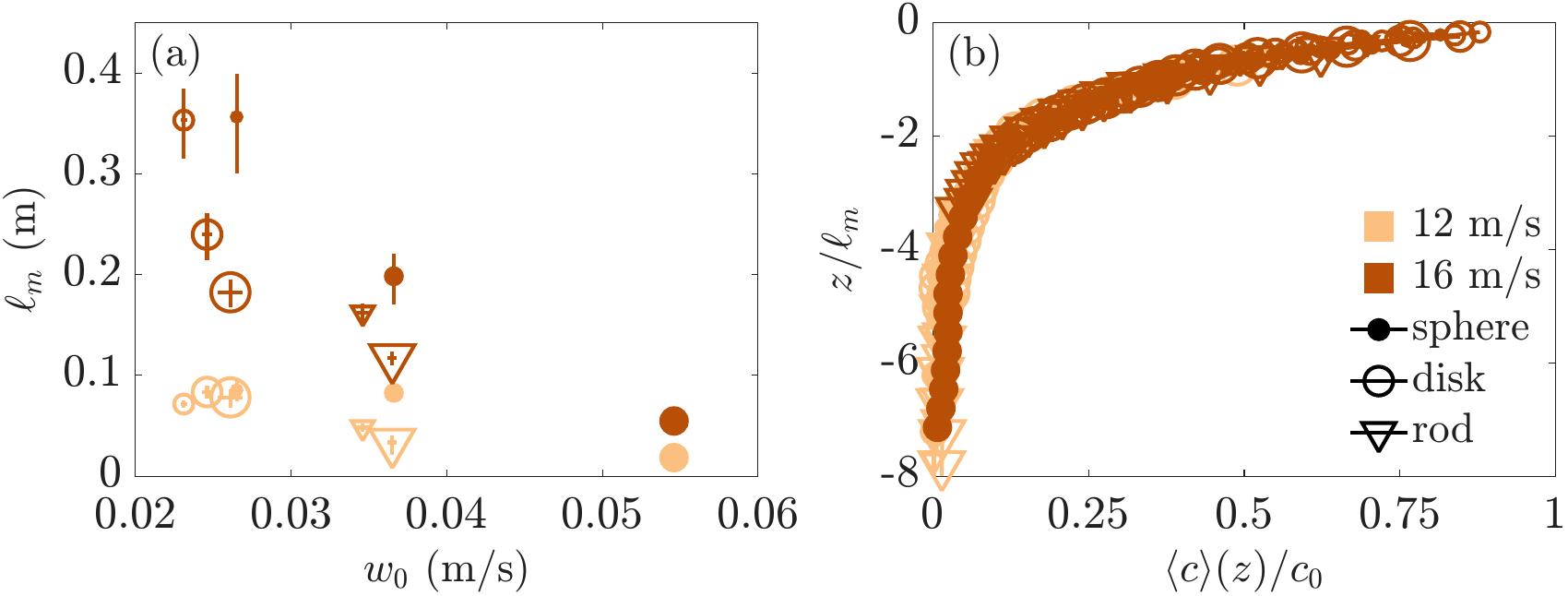}
    \caption{Mixing length and particle concentration profiles derived from exponential fits. (a) Mixing length $\mixlen$ as a function of particle rise velocity $\wvelq$ for all buoyant particle types and wind speeds. Error bars denote $95\%$ confidence intervals from exponential-profile fits. (b) Normalized concentration profiles $\concavg (\zDir)/\concInit$ plotted against normalized depth $\zDir/\mixlen$. }
    \label{fig:exp_norm_Lm_wr}
\end{figure}

\begin{figure}
    \centering
    \includegraphics[scale=0.43]{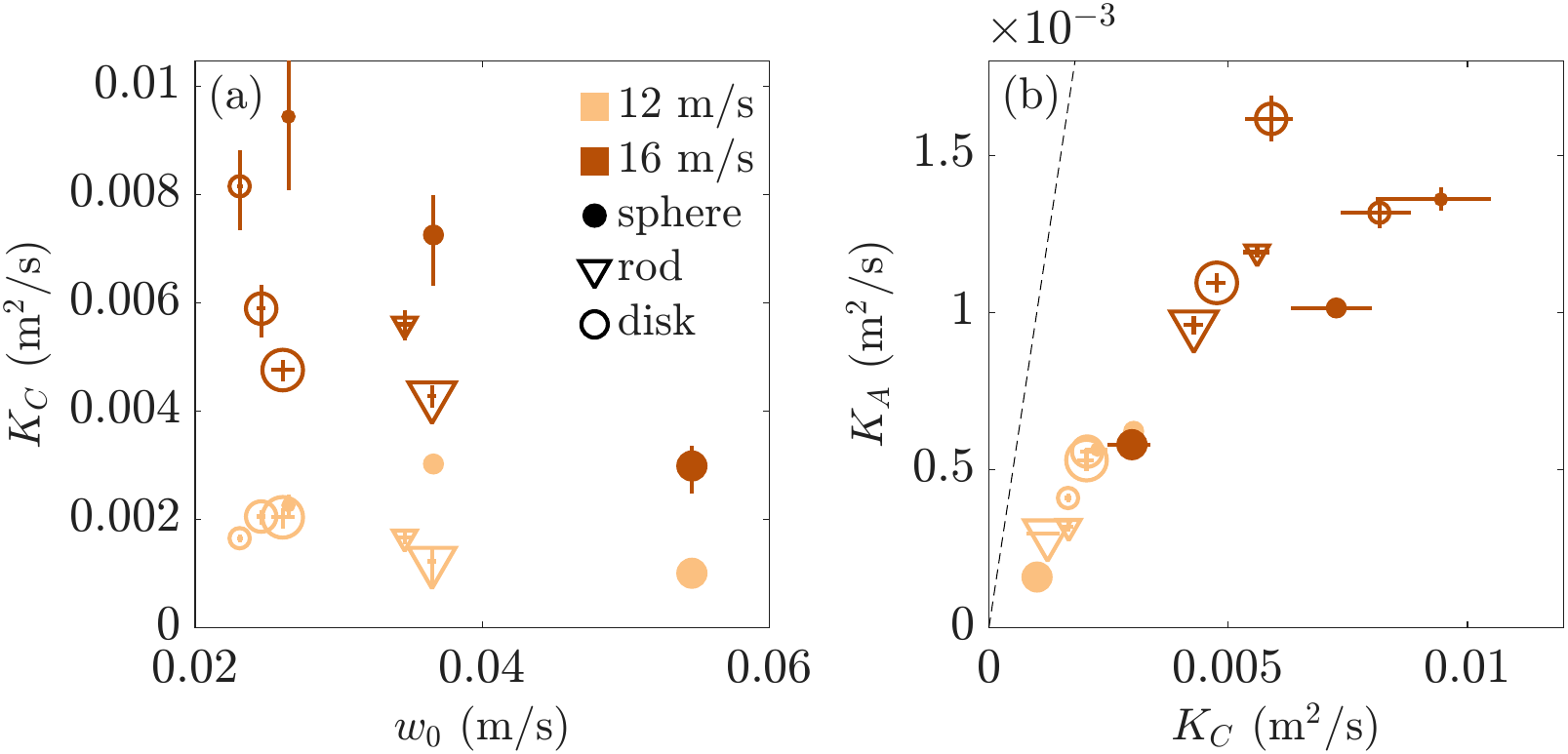}
    \caption{(a) Depth-averaged vertical diffusivity from exponential fitting, $\diffkze = \mixlen \wvelq$ as a function of particle rise velocity. Error bars denote 95\% confidence intervals. 
    (b) Comparison of dispersion-based diffusivity vs. diffusivity from exponential fitting; the dashed line is 1:1.}
    \label{fig:Kz_conc_flux}
\end{figure}

We estimate the average diffusivity from the full concentration profiles. 
In Figure~\ref{fig:conc_prof_normint}, we see an exponential decay of $\concavg$ with depth for all buoyant particle types and wind speeds, agreeing with the predicted profile for a constant diffusivity as in \eqref{eq:vertDistributionModelKukulka}. 
Fitting the function $\concavg (\zDir) = \concInit \exp(\zDir/\mixlen)$ to each profile, we obtain the particle surface concentration $\concInit$ and the mixing length $\mixlen$. 
We plot the mixing length $\mixlen$ as a function of $\wvelq$ in Figure~\ref{fig:exp_norm_Lm_wr}a and see that $\mixlen$ decreases with increasing $\wvelq$, i.e., increasingly buoyant particles are found closer to the surface. 
When we normalize the concentration profiles by $\concInit$ and plot them against depth normalized by $\mixlen$ in Figure~\ref{fig:exp_norm_Lm_wr}b,  all the profiles collapse onto a single curve. 
This collapse supports the validity of the model and its assumption of a constant diffusivity (and drift velocity) over depth. From these fits, we find the corresponding diffusivity as $\diffkze = \mixlen \, \wvelq$ \eqref{eqn:diff_mixlen}, shown in Figure~\ref{fig:Kz_conc_flux}a.  
Under the high wind speed (16~m/s), $\diffkze$ decreases with increasing $\wvelq$ with some shape-dependence, e.g., falling most rapidly for disks. 
In contrast, the 12~m/s wind speed cases show $\diffkze$ is relatively constant across the particle types with a slight decrease with increasing $\wvelq$.

We plot these Eulerian estimates of $\diffkze$ against the Lagrangian estimates of $\diffkzd$ in Figure~\ref{fig:Kz_conc_flux}b, where we observe that the two estimates are strongly correlated. However, the correlation is not one-to-one; the $\diffkze$ values are approximately a factor of $2\text{--}5$ (and in some cases nearly an order of magnitude) larger than the corresponding $\diffkzd$ values. 
Comparing $\diffkzm$ to $\diffkzd$ would give a similar trend, as $\diffkzm$ values track closely to $\diffkze$. 
The likely explanation for the large difference in magnitude is the assumption of $\wvelq$ as the assumed drift velocity in the derivation of $\diffkze$. 
Because $\diffkze = \mixlen \wvelq$ yields diffusivities that are too high, it suggests that $\wvelq$ is an overestimate of the effective particle rise velocity in our experiments. 

\subsection{Effective rise velocity}
\label{subsec:effrisevel}

\begin{figure}
    \centering
    \includegraphics[scale=0.43]{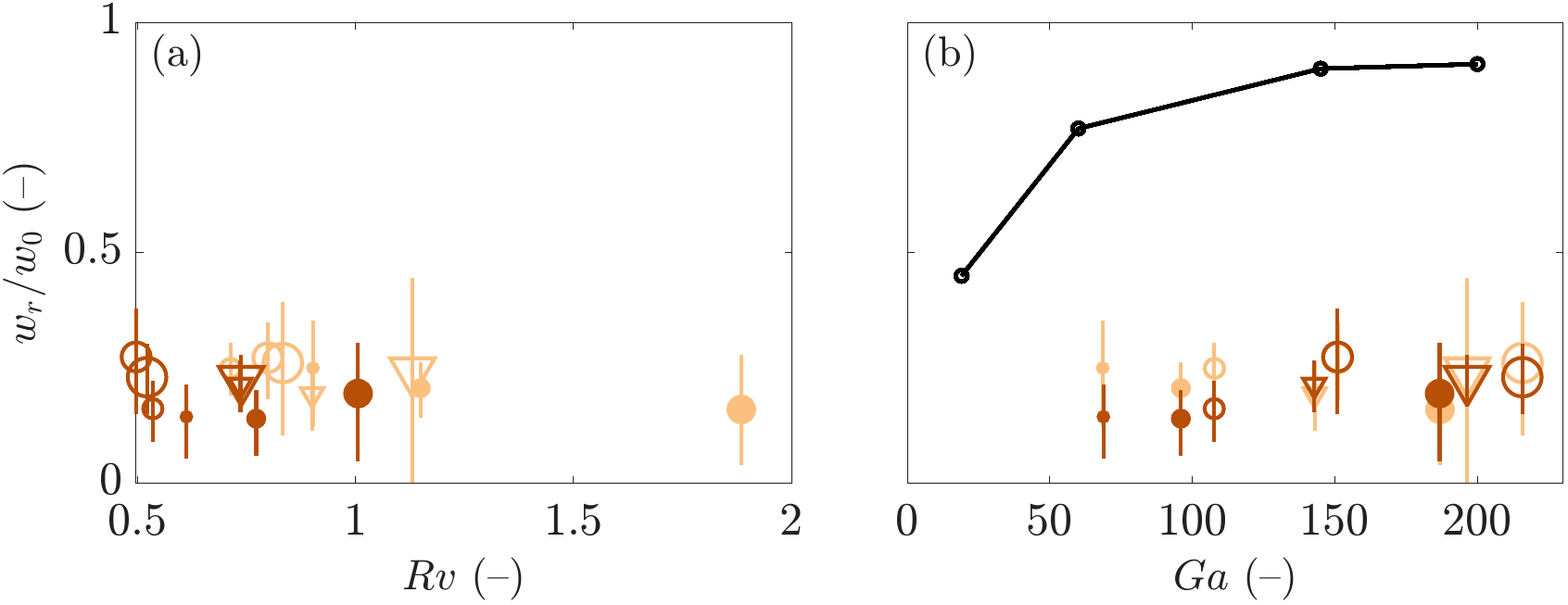}
    \caption{Normalized effective rise velocity $\riseVel / \wvelq$, where $\riseVel = \diffkzd/\mixlen$, versus (a) velocity parameter $\Sv$, and (b) Galileo number $\Ga$ where the black solid line represents the effective rise velocity of settling spheres in HIT from \citet{Fornari_2016a}. The error bars come from error propagation. We omitted the error bars for $\Sv$ and $\Ga$ to improve visualization (see Table~\ref{tab:summaryparticlesflow} for the values with uncertainties).}
    \label{fig:effwq}
\end{figure}

Given that $\diffkze$ overestimates the measured $\diffkzd$ values, we conclude $\wvelq$ is not the appropriate drift velocity to use in this system. 
We instead estimate an effective rise velocity as $\riseVel = \diffkzd/\mixlen$, i.e., the drift velocity that closes the flux balance in \eqref{eqn:diff_flux} given the measured diffusivity $\diffkzd$.

We plot $\riseVel$ normalized by $\wvelq$ as a function of $\Sv$ and $\Ga$ in Figure~\ref{fig:effwq}. 
We find that the effective rise velocity in each case is reduced by up to $80\%$ relative to the quiescent value. 
In Figure~\ref{fig:effwq}b, we compare our results with simulations of settling spheres ($\SG = 1.00035\text{--}1.038$) in homogeneous isotropic turbulence (HIT) \citep{Fornari_2016a}; we justify this comparison by showing that the dynamics of slightly heavy and slightly buoyant particles are approximately symmetric (Appendix~\ref{app:mapping}). 
The reduction we observe far exceeds that observed in HIT at similar $\Ga$.

The excess rise velocity reduction in our experiments is likely due to particles  preferentially sampling downwelling fluid. 
In a flow with a free surface, buoyant particles collect at surface convergence zones which coincide with regions of downwelling; the particles therefore oversample downward-moving fluid, i.e., $\langle w_f \rangle < 0$ along their trajectories. 
This mechanism is well-documented for buoyant material in Langmuir turbulence, where particles accumulate in downwelling regions beneath windrows \citep{Chor_2018, Chamecki_2019}. 
This is similar to the retention of slowly rising particles in cellular flows described in the classic work of \citet{Stommel_1949}. 
While our Langmuir number indicates wave-influenced turbulence without fully developed Langmuir circulation (\S \ref{subsec:flow}), any free-surface turbulent flow can lead to particles clustering in convergence zones \citep{Li_2025}. 

Beyond preferential sampling, some of the rise velocity reduction could be due to the particles' finite size. 
Recent work has argued that the Basset history force and added mass can reduce sediment settling velocities for particles not small relative to the Kolmogorov lengthscale \citep{Li_2023}. 
Our particles have average size ratios of $Sz\approx 9-22$, and the analysis of \citet{Li_2023} predicts a reduction in terminal velocity of up to 65\% for $\Sz=10$ and up to 75\% for $\Sz=20$  (see their Eqn. 10). 
While this model was derived for spherical particles, it does indicate that a large rise velocity reduction is possible for particles that are larger than the Kolomogorov lengthscale.
Because we do not measure the fluid velocity along particle trajectories, our measurements cannot observe or distinguish these effects directly. And, regardless of how the rise velocity is reduced, the implication is the same: namely that the
drift velocity needed to describe the average particle concentration profile  is substantially smaller than $\wvelq$. Conversely, assuming $\wvelq$ would lead to an overestimated diffusivity from the concentration profiles alone.

\subsection{Eddy viscosity and Schmidt number}

\begin{figure}
    \centering
    \includegraphics[scale=0.43]{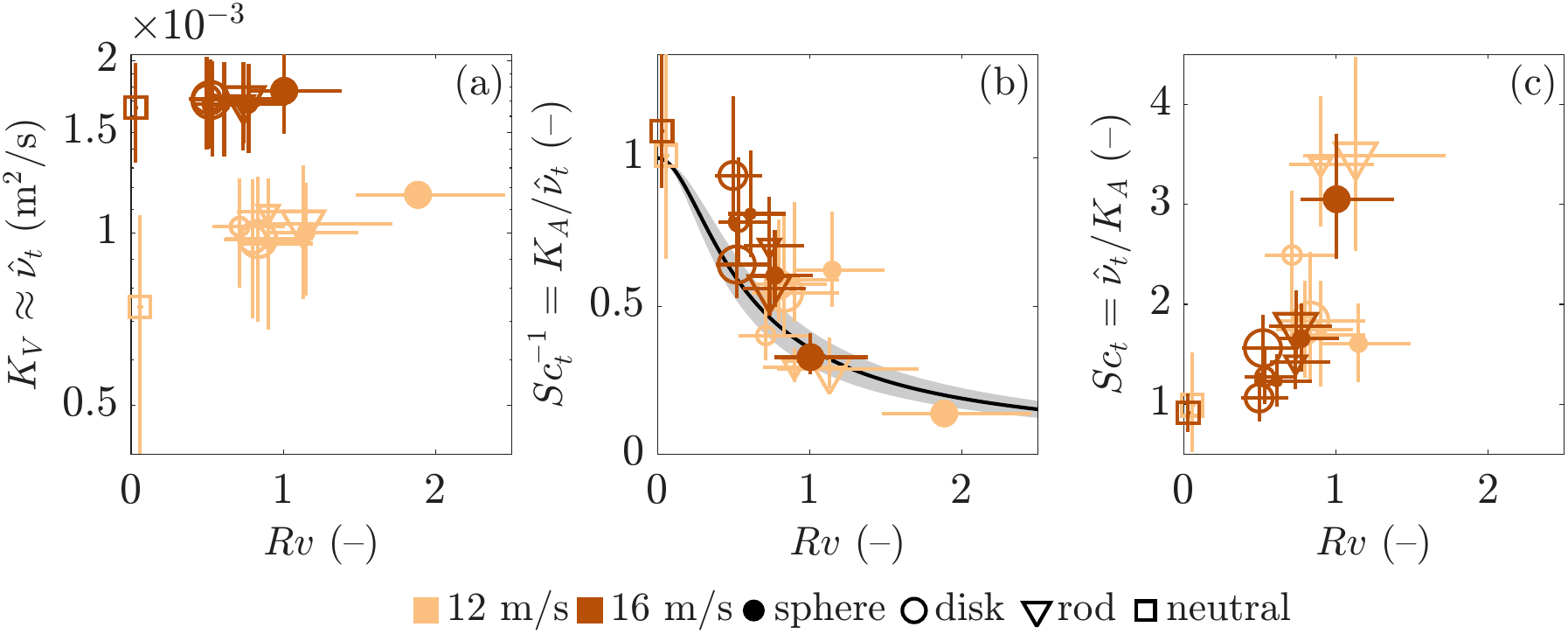}
    \caption{(a) Depth-averaged particle turbulent diffusivity assuming $\Sct \approx 1$ and thus $\diffkzv \approx \eddyVisc$. (b) Inverse turbulent Schmidt number $\Sct^{-1}$ versus $\Sv$ for all particle types and both wind speeds. 
    The solid curve is the \citet{Csanady_1963} crossing-trajectories theory fitting with $\ctrajfac = 2.5$. (c) Turbulent Schmidt number $\Sct$ plotted against $\Sv$.}
    \label{fig:crossTraj}
\end{figure}

Finally, we also compare the particle diffusivity $\diffkzd$ to the flow’s eddy viscosity $\eddyVisc$.
From the eddy viscosity profile in Figure~\ref{fig:dissipgerbi}d, we calculate an effective, average eddy viscosity $\hat{\eddyVisc}$ weighted by the concentration profile for each particle under each wind speed. 
We plot $\hat{\eddyVisc}$ in Figure~\ref{fig:crossTraj}a as a function of $\Sv$, where we can also naively assume $\Sct \approx 1$, and thus $\eddyVisc \approx \diffkzv$. 
We see that $\diffkzv$ increases as $\Sv$ increases, since the more buoyant particles spend increasingly more time close to the surface where $\hat{\eddyVisc}$ is slightly higher.
Although there is only a modest change with $\Sv$, this is still the opposite of the trend observed for $\diffkzd$ and $\diffkze$ with $\Sv$ as shown above.

When we compare the particle and momentum diffusivity as $\Sct^{-1} = \diffkzd/\eddyVisc$ in Figure~\ref{fig:crossTraj}b, we see that $\Sct^{-1}$ decreases as particle buoyancy increases, again consistent with the crossing-trajectories theory of \citet{Csanady_1963} (solid curve from equation \ref{eqn:crosstraj} with best-fit $\ctrajfac = 2.5 \pm 0.5$). 
Consequently, $\Sct$ increases monotonically with $\Sv$ (Figure \ref{fig:crossTraj}c), reaching values up to 3.5 for the most buoyant particles. 
We find that all our buoyant particles have $\Sct > 1$, i.e.,   their diffusivity is significantly less than that of the fluid momentum. 
This result  aligns with that of \citet{Chauchat_2022} who reported $\Sct$ between 3 and 4 for heavy particles in open-channel flows. 
In contrast, the neutral particles exhibit $\Sct\approx1$, indicating no crossing-trajectories effect. 
In addition, this result also indicates strong agreement between the two independent measures of tracer diffusivity: the eddy viscosity from the turbulent Reynolds stresses and the estimate from the neutral particle dispersion.  
Together, these results show that common modeling assumptions (e.g., a quiescent rise velocity and $\Sct=1$) will fail with increasing $\Sv$ for buoyant particles in this system. 
\section{Conclusions}
\label{sec:Conclusions}

This  study characterizes the vertical transport of near-neutrally buoyant, finite-size particles in a wind-driven, free-surface boundary layer, using both Lagrangian and Eulerian approaches to evaluate the particles' vertical turbulent diffusivity $\diffk$. 
The Lagrangian approach estimates  diffusivity $\diffkzd$ directly from the particle dispersion, which follows the classic ballistic-to-diffusive transition, with waves introducing some periodic motion.
A main result is that particle relative rise velocity is the primary control on the vertical diffusivity: $\diffkzd$ decreases by a factor of up to four between the least buoyant and the most buoyant particles, and
this reduction is found to be well explained by the crossing-trajectories theory of \citet{Csanady_1963} across all particle shapes and sizes. 
The effect is further confirmed by the decrease of the fitted Lagrangian decorrelation time scale $\lagtsc$ with $\Sv$.
We also find that particle shape and size primarily are important here through their effect on the quiescent rise velocity $\wvelq$. 

We then show that the Eulerian approach to inferring $\diffkze$ from concentration profiles with an assumed quiescent rise velocity $\wvelq$ overestimates the diffusivity by nearly an order of magnitude compared to $\diffkzd$. 
This implies an effective rise velocity that is reduced up to $80\%$ compared to $\wvelq$, a reduction that exceeds that of settling spheres in homogeneous isotropic turbulence at similar Galileo numbers \citep{Fornari_2016a}. As discussed in \ref{subsec:effrisevel}, this  excess reduction is likely associated with  preferential sampling of downwelling fluid and/or unsteady forces associated with the particles' finite size. To fully characterize the physical mechanisms responsible for the reduced rise velocity, co-located fluid and particle velocities measurements are required.

Finally, we find $\Sct\approx1$ for the neutrally buoyant particles but $\Sct \neq 1$ for the buoyant particles.  
We  also observe that $\Sct$ increases monotonically with $\Sv$, reaching values of up to $3.5$ for the most buoyant particles, values which agree with recent measurements of settling particles in open-channel flows \citep{Chauchat_2022}. 
Thus, this further emphasizes that $\Sct = 1$ is not an appropriate closure for this regime and indicates that these biases are ubiquitous phenomena rather than results unique to a particular flow. Overall, this work highlights the value of Lagrangian measurements, which can both measure particle diffusivity directly and reveal biases that concentration profiles alone cannot. 

\backsection[Acknowledgements]{We gratefully acknowledge Aaron Maschhoff, Inessa Garrey and Anusha Aggarwal for their assistance in the laboratory experiments.}

\backsection[Funding]{This work was supported by the U.S. National Science Foundation (NSF), Grant No. 2237550 and 2537068. L.J.B. was additionally funded by  NSF Grant No. 2126193, and J.E.C.D. acknowledges support from the Link Foundation Ocean Engineering and Instrumentation Ph.D. Fellowship.}

\backsection[Declaration of interests]{The authors report no conflict of interest.}
\appendix

\section{Statistics}\label{app:stats}

\subsection{Assessment of steadiness and streamwise homogeneity in our experiments}
\label{app:steady}
\begin{figure}
    \centering
    \includegraphics[scale=0.38]{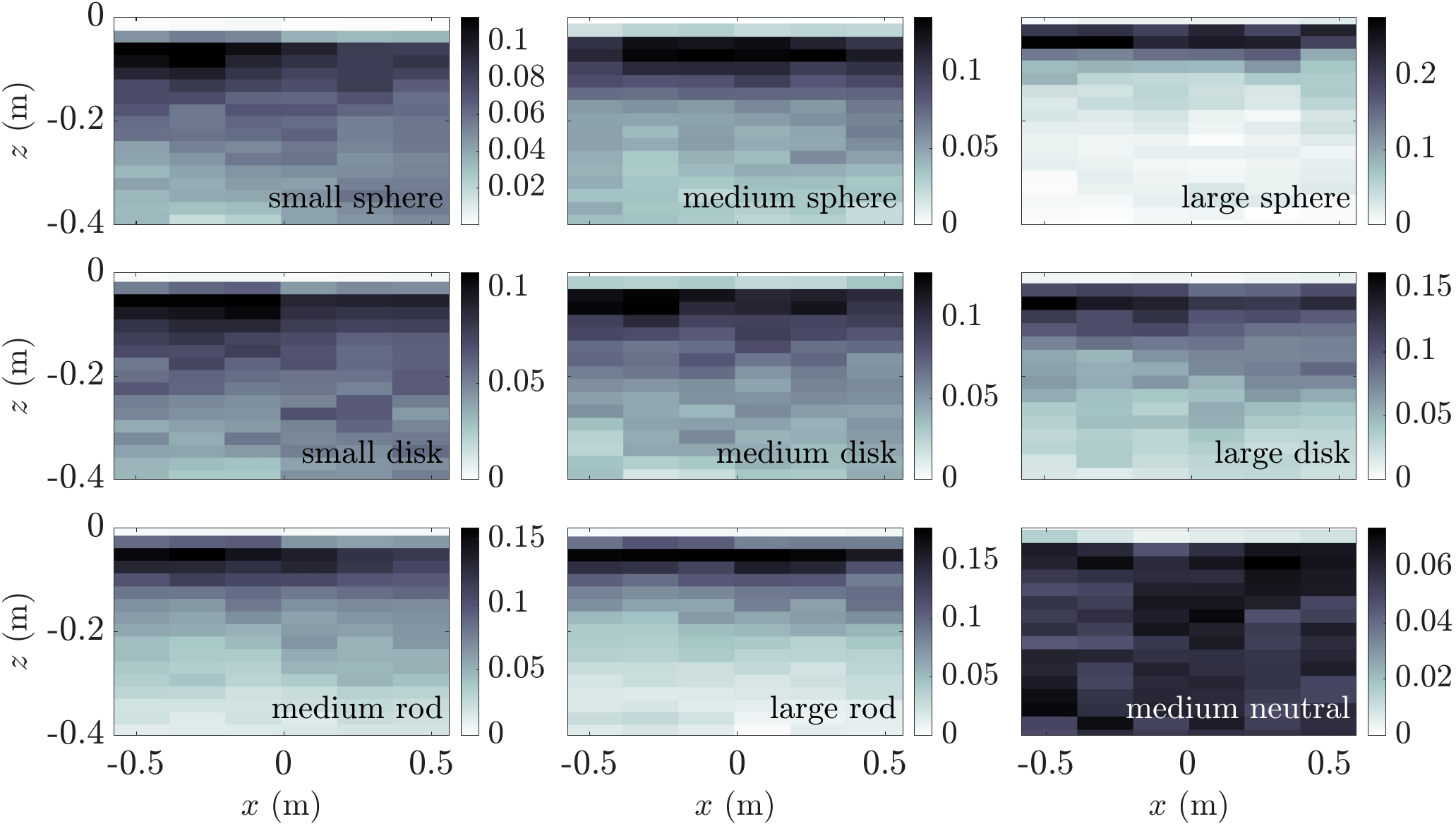}
    \caption{Horizontal-vertical distribution of particle counts (PDF) for various particle types under 16 m/s wind forcing. Each panel displays the probability density of particle counts in the $\xDir\text{--}\zDir$ plane, with darker colors indicating higher concentrations.}
    \label{fig:check_xhomogen16}
\end{figure}

We first check whether the particle distribution changes with horizontal position.
For each particle case, we bin particle positions in the horizontal (15 cm bins) and vertical (3 cm bins) directions, and normalize each horizontal slice by its total particle count. 
We plot the resulting distributions in figure \ref{fig:check_xhomogen16} which shows the buoyant particle concentrations are highest near the free surface and decrease with depth while the neutral particles are more evenly distributed throughout the water column, as expected.
Most particles, including the neutral particles, are mostly homogeneous in the $\xDir$ direction.

\begin{figure}
    \centering
    \includegraphics[scale=0.38]{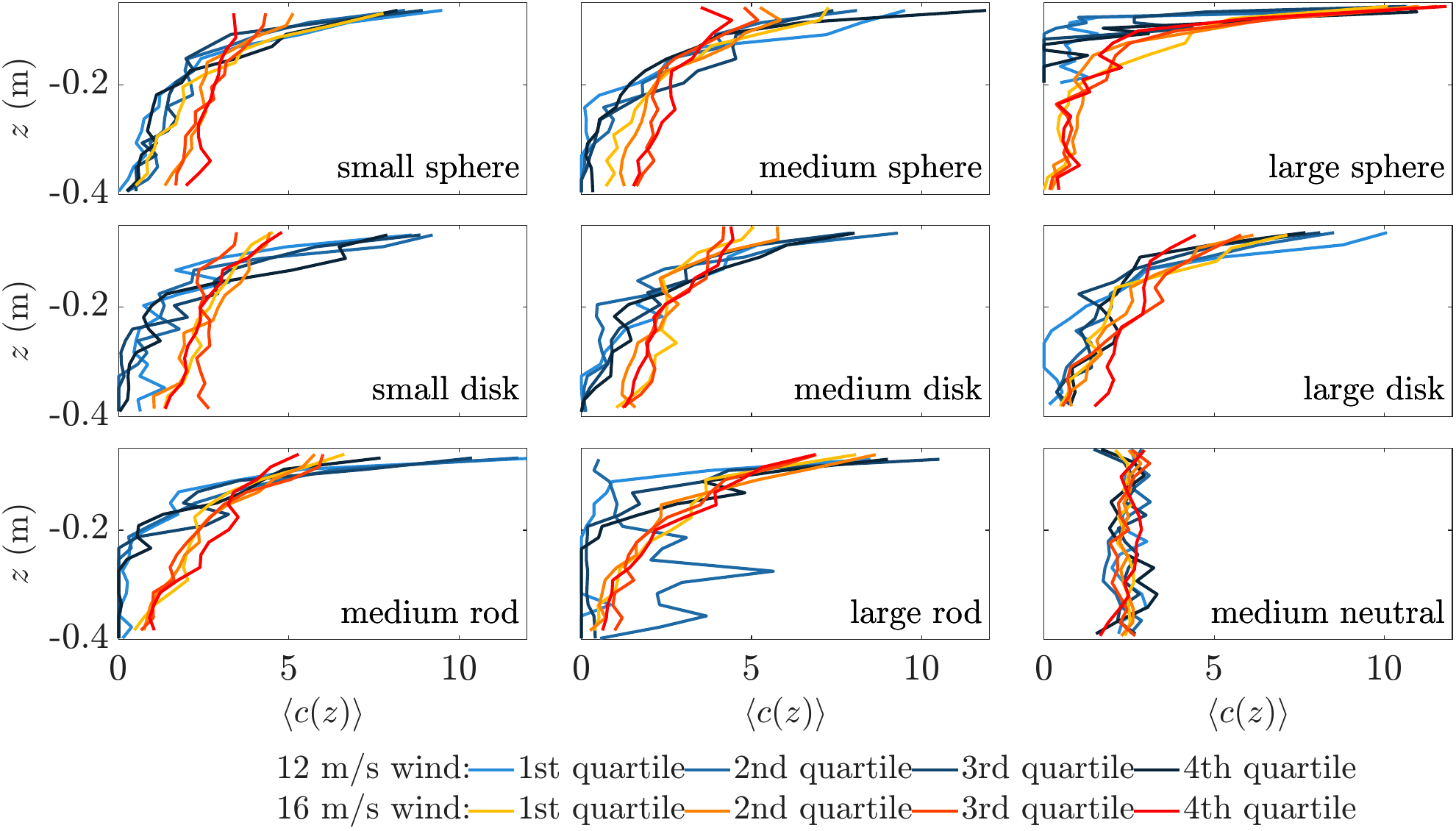}
    \caption{Temporal evolution of vertical concentration profiles $\concavg$ for various particle types under 12 m/s (blue shades) and 16 m/s (red shades) wind speeds.}
    \label{fig:check_steady}
\end{figure}

We then check whether the particles have reached a statistically steady state in figure \ref{fig:check_steady}.
Each total run time was divided into quartiles, and the mean concentration profile for each particle shape, size, and wind speed was computed independently for each quartile. 
The convergence of these four quartile profiles shows that the particle distribution has reached a steady state.
Temporal convergence generally improved with sphere size, though small spheres exhibited persistent transients at 12 m/s. 
Disks and medium rods converged consistently well. 
Large rods were unsteady at 12 m/s, showing pronounced deep penetration, but stabilized at 16 m/s. 
Neutral particles remained stationary and evenly mixed at both wind speeds.
Unsteady quartiles were not considered in the analysis.

\subsection{Trajectory bias correction for asymptotic diffusivity estimate}
\label{app:biascorr}

The raw MSD is biased at large lag because short trajectories do not contribute uniformly across depth. We correct for this by reweighting trajectory segments to match the stationary depth distribution.

We discretize the vertical domain into fine reference bins $\bin=1,\dots,B$ with edges $[\zDir_\bin, \zDir_{\bin+1}]$ and width $\Delta \bin = \zDir_{\bin+1}-\zDir_\bin$. 
From all recorded particle positions, we estimate the stationary depth distribution $\conc(\bin)$ satisfying $\sum_{\bin=1}^{B} \conc(\bin)\,\Delta \bin = 1$.
Then, for a given lag $\tlag$, define the set of all valid start indices $\tlagor$ as $\mathcal S_{\prt}(\tlag) = \left\{ \tlagor : \tlagor+\tlag \le T_\prt \right\}$. The count of valid starts falling in reference bin $\bin$ is
\begin{equation}
\mathcal{C}(\tlag,\bin)
=
\sum_{\prt}
\left|
\left\{
s_0 \in \mathcal S_{\prt}(\tlag) : \zplagr^{(\prt)}(\tlagor)\in \bin
\right\}
\right|.
\end{equation}
with total $N(\tlag)=\sum_{\bin=1}^{B} \mathcal{C}(\tlag,\bin)$ and empirical distribution $\conc_{\tau}(\tlag,\bin) = \mathcal{C}(\tlag,\bin)\,/\,[N(\tlag)\,\Delta \bin]$.
Thus, we define the depth-bias weight as
\begin{equation}
W(\tlag,\bin)
=
\frac{\conc(\bin)}{\conc_{\tau}(\tlag,\bin)},
\end{equation}
which up-weights underrepresented starting depths $\tlagor$ and down-weights overrepresented ones.
If we denote the squared displacement associated with $\tlagor$ as
\begin{equation}
\langle \Delta \zplagr^2 \rangle_{\mathrm{raw}}(\tlag,\bin)=\frac{1}{C(\tau,b)}
\sum_{\prt}
\sum_{\tlagor \in \mathcal S_{\prt,\bin}(\tlag)}
\left[\zplagr^{(\prt)}(\tlagor+\tlag)-\zplagr^{(\prt)}(\tlagor)\right]^2,
\end{equation}
then the corrected MSD is
\begin{equation}
    \langle \Delta \zplagr^2 \rangle_{\mathrm{corr}}(\tlag)
= \frac{
\sum_{b=1}^{B} W(\tau,b)\,C(\tau,b)\,\langle \Delta z^2 \rangle_{\mathrm{raw}}(\tau,b)
}{
\sum_{b=1}^{B} W(\tau,b)\,C(\tau,b)
}.
\end{equation}

This yields a single depth-averaged MSD at each lag. The correction uses the full stationary distribution $\conc(\bin)$ across all fine bins, so the resulting diffusivity $\diffkzd$ is a bulk quantity representative of the entire water column, not a depth-resolved estimate. When $\mathcal{C}(\tlag,\bin)=0$ for bins where $\conc(\bin)$ is appreciable, the lag is too large for reliable sampling and is excluded.
\section{Equivalence between diffusivity in conservation equation and MSD}
\label{app:equivalenceMSDconc}
For a point-source release, the mean concentration $\concavg$ is proportional to the position PDF of an ensemble of particles, and its evolution is described by \eqref{eq:fp}.
At $\tm = 0$, the particles are concentrated at the coordinate origin, so we have $\conc (\zDir, 0 ) = \concInit \delta(\zDir)$, with $\concInit$ the initial integrated concentration and $\delta (\zDir)$ the Dirac delta function. 
The first and second moments of $\zplagr$ follow by multiplying \eqref{eq:fp} by $\zDir$ and $\zDir^2$ and integrating over $\zDir$ after dividing out $\concInit$. For the mean, we have
\begin{equation}
   \frac{\mathrm{d}}{\mathrm{d}\tm}\langle \zplagr(\tm)\rangle
= \int \zDir\,\frac{\partial \concavg}{\partial \tm}\,\mathrm{d}\zDir
= \riseVel,
\end{equation}
so $\langle \zplagr(\tm)\rangle = \riseVel \tm$, as before.
In the same manner, the second moment of \eqref{eq:fp} is
\begin{equation}
\int \zDir^2\,\left[ \frac{\partial \concavg}{\partial \tm} + \frac{\partial}{\partial \zDir}\big(\riseVel \concavg \big) \right ] \,\mathrm{d}\zDir
= \int \zDir^2\,\left[ \frac{\partial}{\partial \zDir}\left(\diffk(\tm)\,\frac{\partial \concavg}{\partial \zDir}\right) \right ] \,\mathrm{d}\zDir,
\label{eq:fp1}
\end{equation}
assuming $\concavg$ and its derivatives vanish at the boundaries, giving
\begin{equation}
   \frac{\mathrm{d}}{\mathrm{d}\tm}\langle \zplagr^2(\tm)\rangle - 2 \riseVel \langle \zplagr(\tm)\rangle = 2 \diffk(\tm).
\end{equation}
With $\langle \zplagr(\tm)\rangle = \riseVel \tm$, we obtain
\begin{equation}
\frac{1}{2}\frac{\mathrm{d}}{\mathrm{d}\tm}\langle \zplagr^2(\tm)\rangle - \riseVel^2 \tm = \diffk(\tm).
\label{eq:fp_moment}
\end{equation}
Thus, the Eulerian conservation equation for concentration and the Lagrangian MSD (see equation \ref{eqn:tay17}) are two representations of the same transport process.

\section{Mapping between slightly buoyant and slightly settling particles}\label{app:mapping}
Under a simplified Maxey–Riley equation, particles with small positive (settling) and negative (rising) density differences exhibit leading-order symmetric behavior, primarily differing by a reversal of the effective gravitational forcing direction.
Given that the Maxey-Riley equation is valid for small, low-$\Rep$ spheres and our particles are neither, we only use this equation to reason about scalings and symmetry, not to model the dynamics quantitatively.

For a rigid sphere in a non‑uniform flow, the Maxey–Riley equation \citep{Maxey_1983} including Stokes drag, fluid acceleration, added mass, and gravitational/buoyancy forcing can be written as
\begin{equation}
\frac{\mathrm{d} \up}{\mathrm{d}\tm} = \denpar \frac{\mathrm{D} \ufp}{\mathrm{D}\tm} + (1-\denpar)\gvec -\resrate(\up - \ufp),
\label{eq:MR_simplified}
\end{equation}
where $\up(\tm)$ is the particle velocity and $\ufp(\xp,\tm)$ the fluid velocity at the particle centroid $\xp$.
The material derivative of the fluid velocity $\mathrm{D} \ufp/\mathrm{D}\tm$ accounts for local unsteadiness and convective fluid acceleration.
The gravitational acceleration vector is $\gvec$ and the coordinate system in $\zDir$ is positive upward. 
The particle response rate $\resrate$, density parameter $\denpar$, and gravitational forcing $\gfvec$ are defined as
\refstepcounter{equation}
$$
  \denpar = \frac{3}{2 \left( \frac{\denp}{\denf} + \tfrac12 \right)},
\quad
\gfvec = (1-\denpar)\gvec = \gvec\frac{\frac{\denp}{\denf} - 1}{\frac{\denp}{\denf} + \tfrac12},
\quad
\resrate = \frac{18\visck \, \mathcal{D}(\Rep)}{\partDiam^2\left(\frac{\denp}{\denf} + \tfrac12\right)},
  \eqno{(\theequation{\mathit{a},\mathit{b},\mathit{c}})}\label{eq:alpha_q_def}
$$
Here, $\rtimep = \resrate^{-1}$ is an effective particle response time and quantifies its inertial response to fluid motion and we have assumed spherical particles with an added mass coefficient of $1/2$. 
In the Stokes limit, $\mathcal{D}(\Rep)=1$ and $\rtimep$ reduces to the usual Stokes relaxation time. 
For finite particle Reynolds number, one common parameterization for the nonlinear drag is the Schiller-Naumann function
$\mathcal{D}(\Rep)=
1+0.15\Rep^{0.687}.$
The vector $\gfvec$ is the effective gravitational or buoyancy acceleration.

To examine the near‑neutral limit, we write the density ratio as
\begin{equation}
    \frac{\denp}{\denf} = 1+ \rdendiff,
    \label{eqn:denratio}
\end{equation}
with $\rdendiff = \Delta \denp / \denf$ measuring the relative density difference. 
For a slightly heavy particle, $\rdendiff>0$ and for a slightly buoyant one, $\rdendiff<0$.
Because we are primarily interested in evaluating how particle inertia and buoyancy respond to small density contrasts, we perform asymptotic expansions on the particle response rate $\resrate$ and the gravitational forcing $\gfvec$. 
Note that $\beta$ and $\resrate$ have the same density-ratio dependence.

Substituting \eqref{eqn:denratio} into $\resrate$ and $\gfvec$ \eqref{eq:alpha_q_def}, and expanding in powers of $\rdendiff$ gives
\refstepcounter{equation}
$$
  \resrate(\rdendiff, \Rep) = \frac{12\visck \mathcal{D}(\Rep)}{\partDiam^2}\left( 1- \frac{2}{3}\rdendiff + O(\rdendiff^2) \right), \text{ and } \gfvec(\rdendiff) = \frac{2}{3} \gvec \rdendiff \left(1 -\frac{2}{3}\rdendiff + O(\rdendiff^2) \right).
  \eqno{(\theequation{\mathit{a},\mathit{b}})}\label{eqn:resrate}
$$

Thus, for fixed $\Rep$, the response rate varies approximately linearly with $\rdendiff$, while its dependence on particle Reynolds number is prescribed through the nonlinear drag function $\mathcal{D}(\Rep)$. The effective gravitational forcing varies smoothly and approximately linearly with $\rdendiff$.
In this asymptotic analysis, $\Rep$ is held fixed, corresponding to a directly measured particle-fluid relative velocity. The factor $\mathcal{D}(\Rep)$ is therefore treated as constant with respect to $\rdendiff$.
We therefore remove the $\Rep$ dependence in the subsequent notation.

Now, consider two particles with the same diameter $\partDiam$ and small density contrast $|\rdendiff| \ll 1$, one slightly heavier than the fluid ($\rdendiff = + \rdendiff_1$ where  $\rdendiff_1>0$) and one slightly lighter ($\rdendiff = - \rdendiff_1$). 
From \eqref{eqn:resrate}, the difference magnitude is for the particle response rate and buoyancy term are
\refstepcounter{equation}
$$
  \resrate(+\rdendiff_1) - \resrate(-\rdendiff_1) = - \frac{16\visck \mathcal{D}(\Rep)}{\partDiam^2} \rdendiff_1 + O(\rdendiff_1^2), \, \text{and} \, |\gfvec(+\rdendiff_1)| - |\gfvec(-\rdendiff_1)| = -\frac{8}{9}\gmag \rdendiff_1^2  + O(\rdendiff_1^3),
  \eqno{(\theequation{\mathit{a},\mathit{b}})}\label{eqn:resratediff}
$$
respectively.
Thus, slightly heavy and slightly buoyant particles with the same diameter experience gravity/buoyancy forcing of opposite sign and equal magnitude up to first order in $\rdendiff_1$, while their response rates agree only to zeroth order.

Their dynamics differ primarily by a reversal of the vertical direction.
Mathematically, if we reflect the vertical variables
\begin{equation*}
    \zDir \mapsto - \zDir, \quad \wplagr \mapsto - \wplagr, \quad \wflagr \mapsto - \wflagr, \quad \gvec\mapsto - \gvec,
\end{equation*}
then, to leading order in $\rdendiff_1$, the equation of motion \ref{eq:MR_simplified} for a slightly buoyant particle is mapped into that of a slightly sinking particle with the same $\rdendiff_1$.
This justifies using results from the extensive literature on settling particle to interpret the behavior of nearly neutrally buoyant particles, provided $|\denp/\denf-1|\ll 1$.
For our particles, $|\denp/\denf-1|\leq 0.04$, and therefore the asymmetry between a buoyant and sinking particle's $\resrate$ and $\gfvec$ values, relative to their leading order values, is $4/3\rdendiff_1\approx 5\%$ for $\rdendiff_1=0.04$.
Also, for these values, the near‑neutral expansion is accurate to approximately $1\%$.

To summarize, the vertical reflection of the dynamics represents a leading-order symmetry of the reduced particle equation when particles have the same diameter, equal density contrast relative to the fluid, and same particle Reynolds number. 
However, this symmetry is approximate since mean shear, free-surface boundary condition, and wave-turbulence dynamics can all distinguish rising from settling particles in wall-bounded or homogeneous flows. The comparison nevertheless provides a useful scaling argument for interpreting and comparing settling and rising near-neutrally buoyant particles results from the literature, provided that the small-density-contrast assumption is satisfied.

\bibliographystyle{jfm}
\bibliography{A_vertDisp}

\end{document}